\documentclass{elsart}
\usepackage{graphicx,amssymb}
\def\fig#1{\includegraphics[width=0.47\hsize]{#1.eps}}
\def\Eq#1{Eq.~(\ref{#1})}
\def\Fig#1{Fig.~\ref{#1}}

\def\tTT{\ifmmode t_{\rm TT}\else$t_{\rm TT}$\fi}
\def\tBDS{\ifmmode t_{\rm BDS}\else$t_{\rm BDS}$\fi}
\def\NEAS{\ifmmode N_\mathrm{EAS}\else $N_\mathrm{EAS}$\fi}
\newcommand{\Padj}{\ensuremath{\mathcal{P}_\mathrm{adj}}}
\def\norm#1{\left\|#1\right\|}
\def\abs#1{\left| #1 \right|}
\def\const{{\rm const}}
\def\art#1#2#3#4#5{#1,   #2   #3 (#4) #5.}
\def\artx#1#2#3#4#5#6{#1, #2 #3 (#4) #5; #6.}
\def\arx#1#2{#1, #2.}
\def\proc#1#2#3#4#5{#1, #2, vol.~#4, (#3) p.~#5.}
\def\book#1#2#3{#1, #2 (#3).}
\def\ApJ{Astrophys.\ J.}
\def\PRA{Phys.\ Rev.~A}

\def\PRE{Phys.\ Rev.~E}
\def\PRL{Phys.\ Rev.\ Lett.}
\def\PD{Physica~D}

\def\PLA{Phys.\ Lett.~A}

\def\JOSA{J.~Opt.\ Soc.\ Am.}

\def\NPBP{Nucl.\ Phys.~B (Proc. Suppl.)}
\def\JGR{J.~Geophys.\ Research}
\def\RAN{Izv.\ Ros.\ Akad.\ Nauk, Ser.\ Fiz.}

\begin{document}
\begin{frontmatter}
\title{BURSTS OF EXTENSIVE AIR SHOWERS:
       CHAOS VS.\ STOCHASTICITY}
\author{Yu.A. Fomin\corauthref{cor}},
\corauth[cor]{Corresponding author.}
\ead{fomin@eas.sinp.msu.ru}
\author{G.V. Kulikov},
\ead{kulikov@eas.sinp.msu.ru}
\author{M.Yu.~Zotov}
\ead{zotov@eas.sinp.msu.ru}

\address{Ultrahigh Energy Particles Laboratory,
        D. V. Skobeltsyn Institute of Nuclear Physics,
        M. V. Lomonosov Moscow State University,
        Moscow 119992, Russia}

\begin{abstract}
     Bursts of the count rate of extensive air showers (EAS) lead
     to the appearance of clusters in time series that represent
     EAS arrival times.
     We apply methods of nonlinear time series analysis to~20
     EAS cluster events found in the data set obtained with
     the EAS--1000 prototype array.
     In particular, we use the Grassberger--Procaccia algorithm
     to compute the correlation dimension of the time series in
     the vicinity of the clusters.
     We find that four cluster events produce signs of chaotic
     dynamics in the corresponding time series.
     By applying a number of supplementary methods we assess that the
     nature of the observed phenomenon may indeed be chaotic and thus
	deterministic.
     We suggest a simple qualitative model that might explain an
	origin of EAS clusters in general and
     ``possibly chaotic'' clusters in particular.
     Finally, we compare our conclusions with the results of similar
     investigations performed by the EAS-TOP and LAAS groups.
\end{abstract}

\end{frontmatter}

\section{Introduction}

     We have already studied the distribution of arrival times of
     extensive air showers (EAS) registered with the EAS--1000
     prototype array both by methods of classical statistics
     \cite{Dubna,Hamburg} and by methods of cluster
     analysis \cite{IzvRAN01,clusters'02}.
     In particular, we presented 20 EAS cluster events---groups
     of consecutive showers that were registered in time intervals
     much shorter than expected ones.
	These groups represent bursts of the EAS count rate and
	may be considered as cosmic ray bursts,
	see \cite{IzvRAN01,clusters'02}.
     This phenomenon has put forward at least two questions: one
     on the astrophysical nature of the process, another on its
     statistical properties.
	Namely, while the vast majority of sufficiently long samples
	satisfy the hypothesis for an exponential distribution of time
	delays between EAS arrival times \cite{Dubna,Hamburg}, the
	$\chi^2$-test performed for samples taken in the vicinity of some
	of the EAS clusters made us reject this hypothesis.
     Thus we have decided to apply methods of nonlinear time series 
	analysis to samples that contain EAS clusters in order to clarify 
	dynamical reasons of this situation.
	This approach has already proved to be a powerful tool of
	investigation in different fields of science including different
	fields of astrophysics in general~\cite{CP87&GRL91} and cosmic ray 
	physics in particular, see
	\cite{Aglietta,Bergamasco92,Bergamasco94,Kudela,Japan:AP,Japan2001}
	and references therein.

	Recall that the data set under consideration represents 203~days
	of regular operation of the array for the period August~30,
	1997 to February~1, 1999.
	The total EAS number in the data set equals 1~668~489.
	The mean number of charged particles in a shower is of the order
	of~$1.2\times10^5$.
	This corresponds to the energy of a primary particle 
	$\sim1$~PeV/nucleon.
     The mean interval between consecutive EAS arrival times is
     equal to 10.5~s.
     The discreteness in the moments of EAS registration 
     approximately equals 0.055~s (one tic of the PC clock).

     In~\cite{nonlinear'02}, we have already discussed briefly the
     results of nonlinear analysis of one of the EAS clusters.
	In the present paper, we study this cluster in depth and briefly
	discuss three other clusters that demonstrate nonlinear features.

\section{Preliminary Facts}

     In what follows, we study a scalar time series of the form
     $x_1, x_2,\ldots, x_n$, where $x_i = t_i - t_{i-1}$, $t_i$ is
     the moment of registration of the $i$th EAS, $i=1,\ldots,n$, and
     $t_0=0$ corresponds to midnight.
     For the purposes of our investigation, it is convenient to use
     time intervals~$x_i$ given in seconds.

	One can point out two main approaches to nonlinear analysis
	of the time series: (1)~the scaling exponent and fractal length
	estimates based on the self-similarity properties of
	experimental signals~\cite{Man77}; (2)~an estimate of the correlation
	dimension based on the embedding of the original time series in
	an $m$-dimensional phase space \cite{Packard80,Takens80,Mane}
	(see, e.g., \cite{Schreiber:PhysRep} for a comprehensive review).
     In the current research, we follow the second approach.
	Within it, the main tool to analyse the dynamics of the time series
	is the Grassberger--Procaccia method~\cite{GP83} with the
	modification suggested by Theiler~\cite{Theiler:W}.
     Namely, given a sample $(x_1,x_2,\ldots,x_N)$ extracted from the
     time series we construct $m$-dimensional delay vectors
\[
     {\bf x}_i = (x_i, x_{i+\tau}, x_{i+2\tau},\dots,x_{i+(m-1)\tau}),
\]
     where~$\tau$ is an arbitrary but fixed time increment and~$m$ is
     an embedding dimension.
     Then we compute the number~$K(\rho)$ of vectors with mutual distance
     less or equal than~$\rho$ and such that delay vectors~${\bf x}_i$
	are shifted by at least~$W$ indices:
\begin{equation}
     K(\rho) = \sum _{i=1}^{M-W} \sum _{j=i+W}^{M}
               \Theta(\rho - \norm{{\bf x}_i - {\bf x}_j}),
     \label{eq:K}
\end{equation}
     where $\Theta$ is the Heaviside step function,
     $M=N-(m-1)\tau$, and $W\ge1$ is the cut-off parameter (the
     Theiler window).
     Finally, we plot $\log K(\rho)$ vs.\ $\log\rho$.
     For small~$\rho$, the slope of this plot is an estimate of
     the correlation dimension~$D_2$:
\[
     D_2(\rho) = \frac{\d \log C_2(\rho)}{\d \log \rho},
\]
     where $C_2$ is the correlation sum:
\[
     C_2(\rho) = \frac{2 K(\rho)}{(M - W)(M - W + 1)}.
\]
     (Obviously, one may use $K$ instead of~$C_2$ in the expression
     for the correlation dimension.)
     A plateau observed in the $D_2(\rho)$-plot for small~$\rho$ or,
     equivalently, a so called scaling region in the $\log K$
     vs.~$\log\rho$ plot are regarded as signs of chaotic dynamics
     in the corresponding time series.
     A value of~$D_2$ at the plateau is taken as an estimate of the
     correlation dimension of the attractor underlying the data.
     This quantity also gives a (lower) estimate for the number of
     degrees of freedom in the process under consideration.

	To compute~$K$, we normally divide each unit interval in $\lg\rho$
	into~50 subintervals of equal length.
	We have found that though this number of subintervals seems to be
	small, it adequately reflects the qualitative structure of~$D_2$
	while a much bigger number of subintervals ($\ge200$) leads to
     considerable fluctuations in~$D_2$; these fluctuations hide
     the structure of~$D_2$, especially for~$m$ large enough.
     A smaller number of intervals gives too rough structure of~$D_2$.
     The derivative is calculated via a standard three-points
	algorithm.
     No smoothing or fitting procedures are used.

     Notice that \Eq{eq:K} contains four free parameters: $N$, $\tau$,
     $W$, and~$m$.
     At the preliminary stage of the investigation, we ``scanned"
     the experimental data set having split it into adjacent samples
     with $N=128$, 256, 512, and~1024.
     At this step, we used $\tau=1$, $W=1$, and odd values of~$m$ in
     the range from~5 to~13.
     The value of~$N$ was chosen to be a power of~2 because this
     allows one to use fast algorithms of calculating the Fourier
     transform that is the main part of traditional Fourier analysis.
     After we had found a number of samples that demonstrated some
     kind of plateau in the plot of~$D_2(\rho)$, we studied the
     corresponding data with different~$N$ in the above range and
     different values of~$\tau$ in the range 1--10.
     (Evidently, in our case~$\tau$ may take only integer values.)
     For~$N$ large enough, we have performed calculations for~$m$
     up to~25.
     In any case, $m$ and~$\tau$ were chosen such that the number of
     delay vectors was greater than~100.
     Finally, to avoid autocorrelation in the time series,
     we employed $W=1\dots20$.

	To compute mutual distances between delay vectors  (see \Eq{eq:K}),
	we tried several norms: the maximum norm~$L_\infty$, the taxicab
	norm~$L_1$, and the Euclidian norm~$L_2$.
	Recall that for a given $m$-dimensional vector~${\bf x}$
	these norms are defined in the following way:
\[
     \norm{\bf x}_\infty = \sup _{1\le i\le m} \abs{x_i},
	\qquad
     \norm{\bf x}_1 = \sum _{i=1}^m \abs{x_i},
     \qquad
     \norm{\bf x}_2 = \sqrt{ \sum _{i=1}^m \abs{x_i}^2}.
\]
	Besides these, we have also employed the ``dimension scaled''
	norms $L_{1C}$ and $L_{2C}$ that are expected to provide more
	reliable results than their classical counterparts
	\cite{Frank93,PK98}:
\[
     \norm{\bf x}_{1C} = \frac{1}{m} \norm{\bf x}_1,
     \qquad
     \norm{\bf x}_{2C} = \frac{1}{\sqrt{m}} \norm{\bf x}_2.
\]
     Since the process of calculating~$D_2$ is very time consuming,
     we have used only one of these norms, namely the maximum norm,
     to analyse the complete data set.
     Four other norms were tested for a number of samples both with
	and without a plateau in the~$D_2(\rho)$-plot obtained
     with~$L_\infty$.

     There are a number of tools that can help one to verify the 
	results of calculating the correlation dimension~$D_2$, see, 
	e.g.,~\cite{Schreiber:PhysRep}.
     Among them, we have chosen the Theiler--Takens ``maximum
     likelihood'' estimator of the correlation
     dimension~\cite{Takens85,Theiler88}:
\begin{equation}
     \tTT(\rho) = C_2(\rho)
               \left[
               \int _0^{\rho}{C_2(\rho')\over\rho'} \, d\rho'
               \right]^{-1}.
     \label{eq:tTT}
\end{equation}
	This quantity not only provides an efficient means to find out
	an optimal estimate of the correlation dimension out of the
	correlation integral but also provides an estimate of the
	statistical error.

     Another problem in the case when a plateau in a plot
	of~$D_2(\rho)$ is observed is to make an assessment about
	the nature of the dynamics.
     The main difficulty is to figure out whether one witnesses
     chaotic dynamics in a deterministic process or a special class
     of stochastic processes~\cite{OsbPro,PSVM}.
     One of the main tools to solve this problem is the method of
     surrogate data~\cite{TheilerPrichard96,Theiler92},
	see also~\cite{Schreiber:surrogates} and references therein.
	The main idea of this approach is to generate a sufficient
	number of artificial (``surrogate'') samples that have
	the same statistical distribution (and possibly some other
	features) as the experimental data and to study
	the behaviour of the correlation dimension.
	If surrogate data demonstrate a plateau in plots of~$D_2(\rho)$
	then one concludes that the original data is stochastic;
	otherwise they are likely to be deterministic and chaotic.

     Among other tools one can find the quantity
\begin{equation}
     \tBDS(m,\rho) = \frac{C_2(m,\rho)}{C_2(1,\rho)^m}
     \label{eq:tBDS}
\end{equation}
     suggested in~\cite{BDSL} and in the above form---in~\cite{SS97}.
     As it was shown in~\cite{BDSL}, for a sequence of independent
     random numbers, $C_2(m,\rho)=C_2(1,\rho)^m$ holds, where~$m$ is the
     embedding dimension, and thus $\tBDS(m,\rho)\equiv1$.
	Therefore, if one finds $\tBDS(m,\rho)\ne1$ for~$\rho$ where
	a plateau in the~$D_2(\rho)$-plots is observed then one may
	conclude that the nature of the dynamics is chaotic.

     Besides this, we used a function suggested in~\cite{Schreiber93}:
\begin{equation}
     \phi_0(\rho) = \frac{D_2(m_2, \rho)-D_2(m_1, \rho)}{m_2-m_1},
     \label{eq:varphi0}
\end{equation}
     where $m_1$ and $m_2$ are different embedding dimensions.
     For independent random data, $\phi_0(\rho)\ne0$ and $\phi_0\to1$ as
     $\rho\to0$.
     We also employed the ``normalized slope'' introduced
     in~\cite{PK98}:
\begin{equation}
     \phi(m, \rho) = \frac{1}{m} D_2(m, \rho).
     \label{eq:varphi}
\end{equation}
     If $\phi(m,\rho)$ does not converge to~0 in a wide range of~$\rho$
	as~$m$ grows but to some value $\ge0.1$, then most probably data
	do not represent a chaotic process but should be treated by
	statistical techniques~\cite{PK98}.

     Recall that before applying nonlinear techniques to the analysis of
     a time series, it is strongly suggested to check whether the time
     series is really nonlinear.
     In particular, one can try a measure for time-reversibility, which
	is considered to be a good indicator for nonlinearity~\cite{SS99}.
     For the data sorted in time order,
\begin{equation}
     \gamma = \frac{1}{(\sigma^2)^{\frac{3}{2}}(N-1)}
               \sum _{i=2}^{N}
               \left( \frac{x_i - x_{i-1}}{s_i - s_{i-1}} \right)^3
     \label{eq:gamma}
\end{equation}
	is calculated, which is just the mean of the slopes, taken to the
     third power; here~$s_i$ are the moments of time
	and~$\sigma^2$ is the variance of the sample.
	(In our case, $s_i - s_{i-1} = 1$ since~$x_i$ represent
	just numerated time delays.)
     For a time series generated by a linear process and for
     the surrogate data, one expects $\gamma \approx 0$.
     In contrast, time series with nonlinearities can be asymmetrical in
     time and may yield values of $\gamma \neq 0$.
     To check, whether~$\gamma$ significantly deviates from zero for
     the studied sample, one should generate a sufficient number of
     surrogate data.
     To pay regard to deviations in both directions ($\gamma>0$ and
     $\gamma<0$), a two sided test has to be
     performed~\cite{TheilerPrichard96}.

     Finally, a few words are in order about stationarity of the time
     series under consideration.
     As is well known, a fundamental assumption underlying almost all
     existing linear and nonlinear techniques of time series analysis is
     that the time series is stationary, see,
	e.g.,~\cite{EckmannRuelle1985}.
     As we have already mentioned
     earlier~\cite{Hamburg,clusters'02,SLC},
     the count rate of EAS depends on the value of the atmospheric
     pressure.
     For the data obtained with the EAS--1000 prototype array,
     this dependence can be approximately expressed by a simple formula
\[
     \ln \NEAS = -\beta P + \const,
\]
	where $\NEAS$ is the number of EAS registered in a time unit,
     $P$ is the atmospheric pressure, mm~Hg, and~$\beta\approx10^{-2}$
     is the barometric coefficient.
     This effect makes the time series non-stationary.
     To provide stationarity (at large time scales) we adjusted time
     delays between consecutive showers to the atmospheric pressure
     $P^* = 742$~mm~Hg which is close to the average pressure for the
     whole analyzed data set.
     This adjustment was made by the following formula:
\[
     x_i = x_i^0 \exp [\, \beta(P^* - P_i)],
\]
     where $x_i^0$ is the experimental time delay, and
     $P_i$ is the atmospheric pressure at the $i$th delay.
     The barometric coefficient was chosen to be
     $\beta=1.08\cdot10^{-2}$, exactly as in our previous
     paper~\cite{clusters'02}.
     Due to the adjustment, the mean and the variance of the time
     series at large time scales were approximately constant.

\begin{rem}
     In fact, since we shall discuss only comparatively
     short samples, the above adjustment to~$P^*$ is not important
     for the present results because atmospheric pressure does not
     change significantly during periods of time covered by the
     samples.
     We perform the adjustment in order to guarantee stationarity
     of the whole time series.
\end{rem}

     To compute the correlation sum~$C_2$ effectively, we have worked out
     an algorithm based on preliminary sorting of mutual distances
     between delay vectors.
     Though quite simple, the algorithm occurred to be up to~10 times
     faster than a straightforward computation of~$C_2$.
     To perform calculations, we employed GNU Octave~\cite{Octave}
     running in Mandrake Linux.

\section{The Main Results}

     As is well known, one needs sufficiently long samples to perform
     a successful time series analysis.
     On the other hand, the bursts of the EAS count rate
     have comparatively short time range.
     Thus we did not in fact expect to find signs of chaotic dynamics
     in the vicinity of EAS clusters.
     Surprisingly, we have found some.

     Having ``scanned'' the available data set, we found that
	for the overwhelming majority of data samples no scaling region
	is observed in $\log C_2$ vs.\ $\log\rho$ plots.
     This is quite natural since arrival times of extensive air
     showers represent a simple stochastic process such that for
     sufficiently long samples, the number of EAS registered in a
     time unit obeys the Poisson distribution in a wide range of time
     delays, see, e.g.,~\cite{Hamburg}.
     Still, we have come across a number of samples with a plateau
     in the~$D_2(\rho)$-plot in the vicinity of four EAS clusters,
     namely those registered on May~14, November~11, and December~28,
     1998 and on January~8, 1999.
     Let us begin our discussion with the second of these events,
	which is the longest one.

\begin{figure}
\begin{center}
     \fig{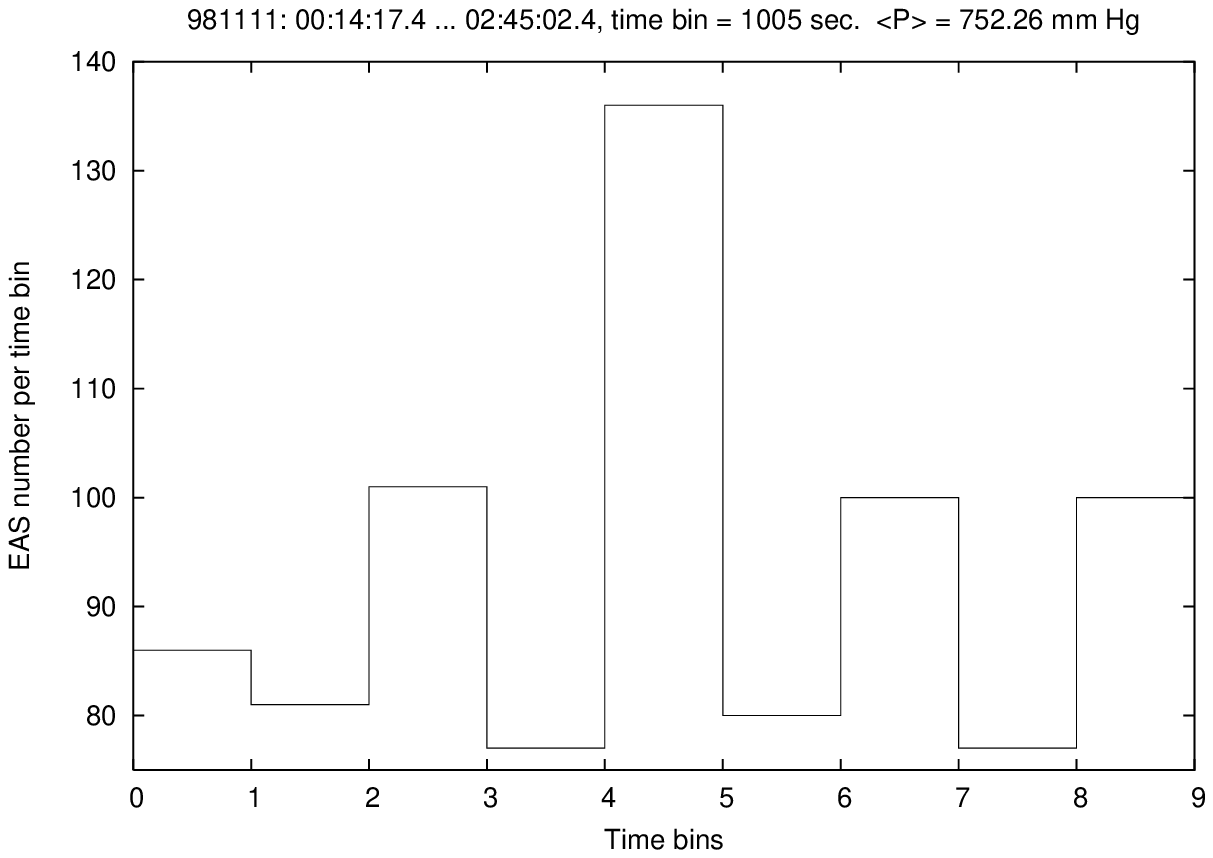}     \fig{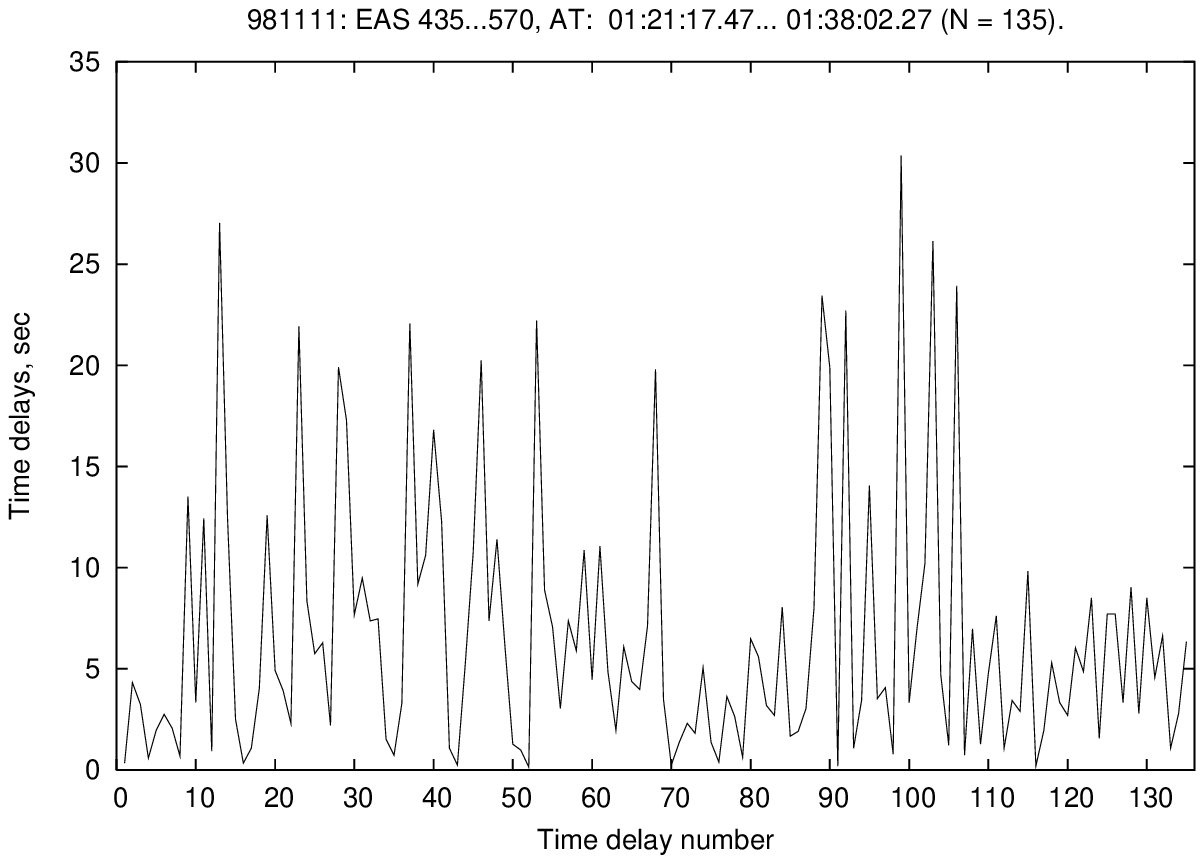}
\end{center}
\caption{The count rate at a time interval that contains the cluster
     registered on November~11, 1998 (left);
     the cluster is presented by the central bin.
     Time delays~(s) between EAS that constitute the cluster (right).
	The delays are adjusted to $P^*=742$~mm~Hg.}
\label{981111:countrate}
\end{figure}

     The cluster event observed on November~11, 1998 consists of 136
     EAS registered within the period from 01:21:17.47 to 01:38:02.27
     (Moscow local time, MSK\footnote{Recall that MSK is connected
	to UT as follows:
	$T_\mathrm{MSK} = T_\mathrm{UT} + 3 + \Delta$ (hours),
	where $\Delta=1$ during Daylight Saving Time, 0~otherwise.})
	with the atmospheric pressure
     $P=752.3$~mm~Hg, see \Fig{981111:countrate}.
     Within the whole data set obtained on November~11, 1998 the showers
     that form the cluster event have numbers 435--570.
     The event is made up of three clusters, which begin at consecutive
     arrival times (i.e., the first cluster begins at the shower~\#435,
     the second one begins at the shower~\#436, etc.)
     and end up simultaneously at the shower~\#570.
     In our opinion, the appearance of three clusters does not reflect
     any process of astrophysical nature but is caused by the technique
     of their selection (see~\cite{clusters'02} for the details).
     Thus we treat the event as a single (outer) cluster.
     The real duration of this cluster equals 1004.8~s while the adjusted
     duration equals 898.8~s; the probability of the appearance of such
     a cluster is of the order of~$2\times10^{-7}$.

     Let us take a look at a number of samples in the vicinity of
	the cluster in order to see how the correlation
     dimension~$D_2$ changes when the cluster appears.
	To make this influence more clear we shall not only present
	samples with a plateau in the plot of~$D_2(\rho)$ but also
	a typical sample outside the cluster.
	Besides nonlinear tools, we shall employ the classical Fourier
	analysis.
     Recall that it is suggested for the Fourier analysis to have
     $x(1) \approx x(N)$.
     All samples discussed below are chosen to satisfy this demand.
     In particular, we omit the last shower of the cluster
     and consider a sample that consists of~134 instead of~135 delays.
     Obviously, this does not influence~$D_2$ and other quantities
	significantly.

\begin{figure}
\begin{center}
	\fig{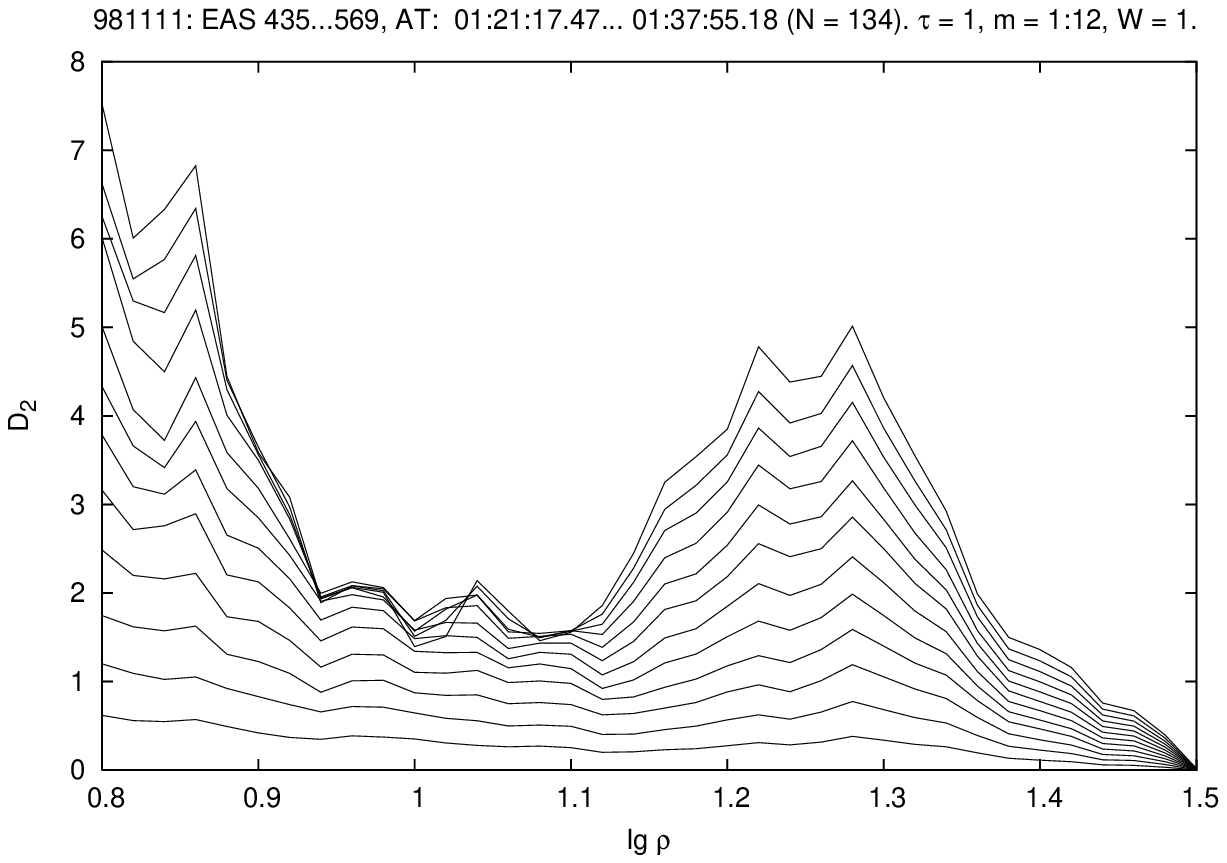}
	\fig{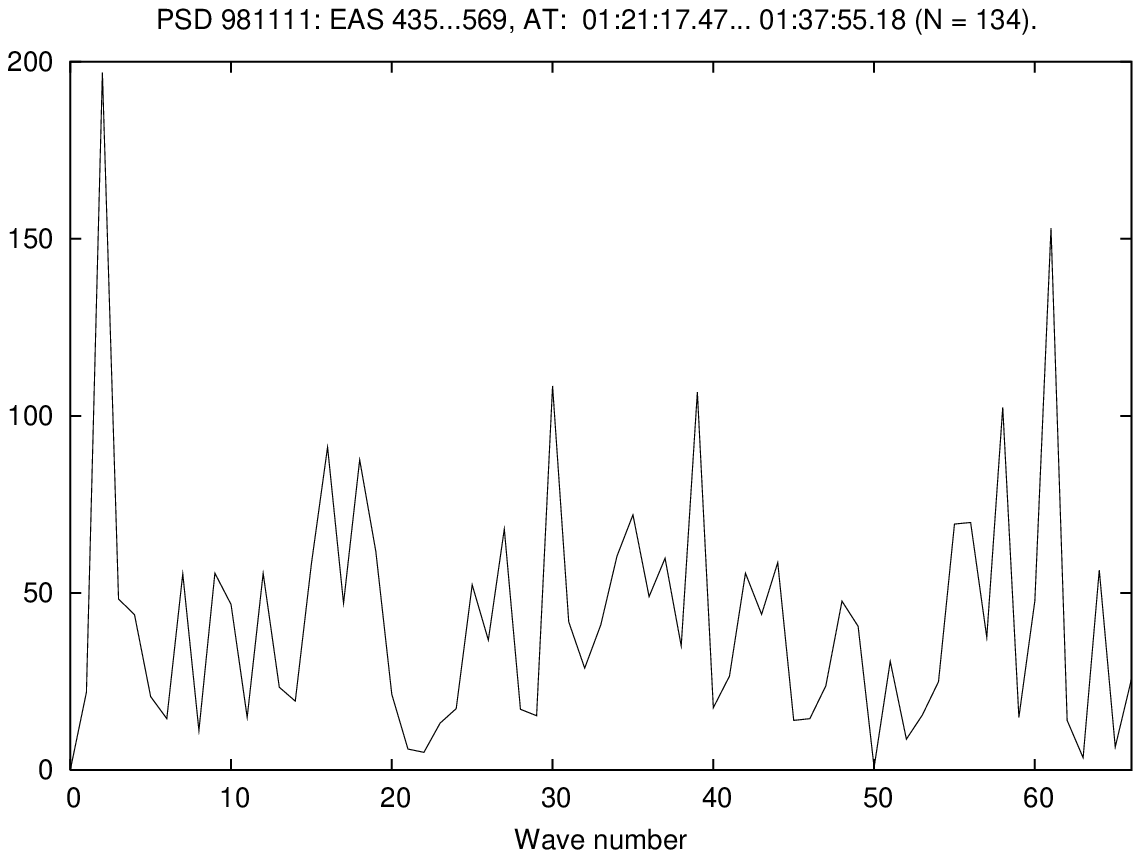}\\
	\fig{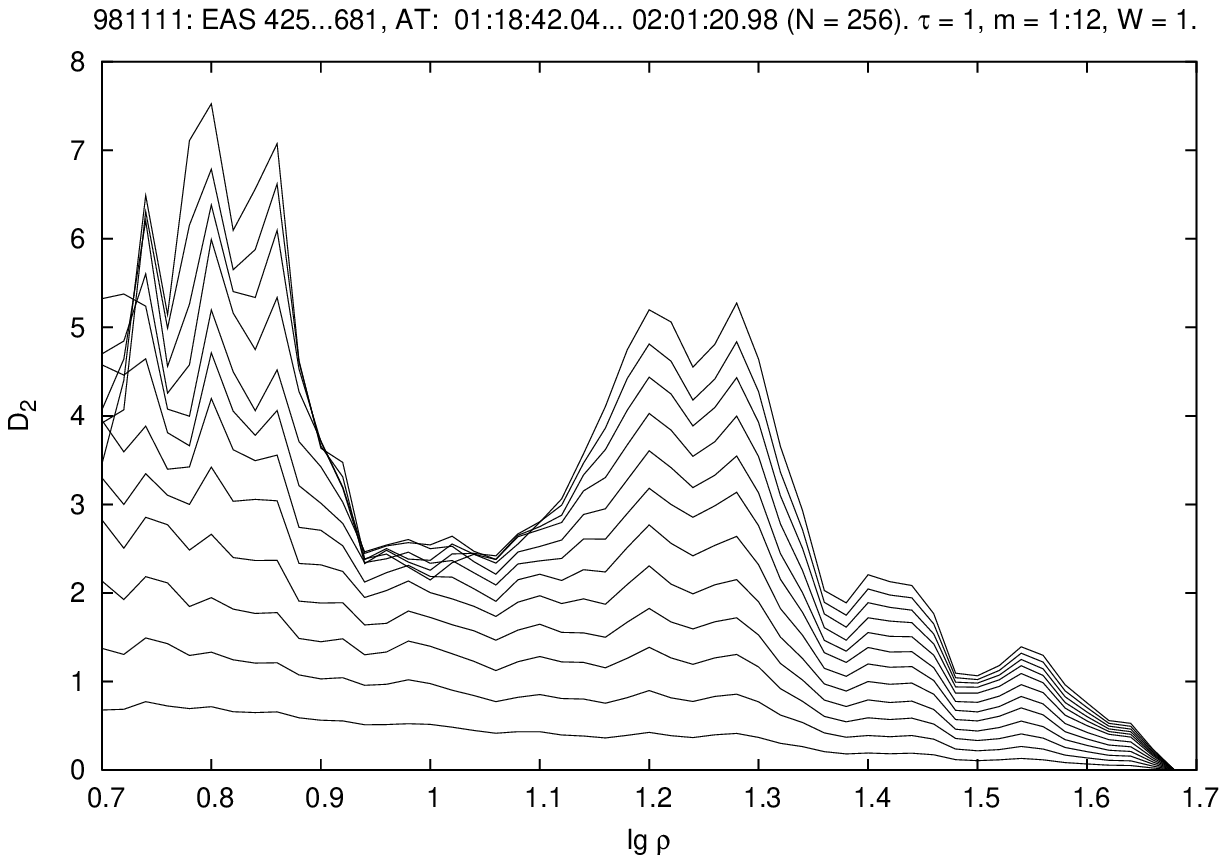}
	\fig{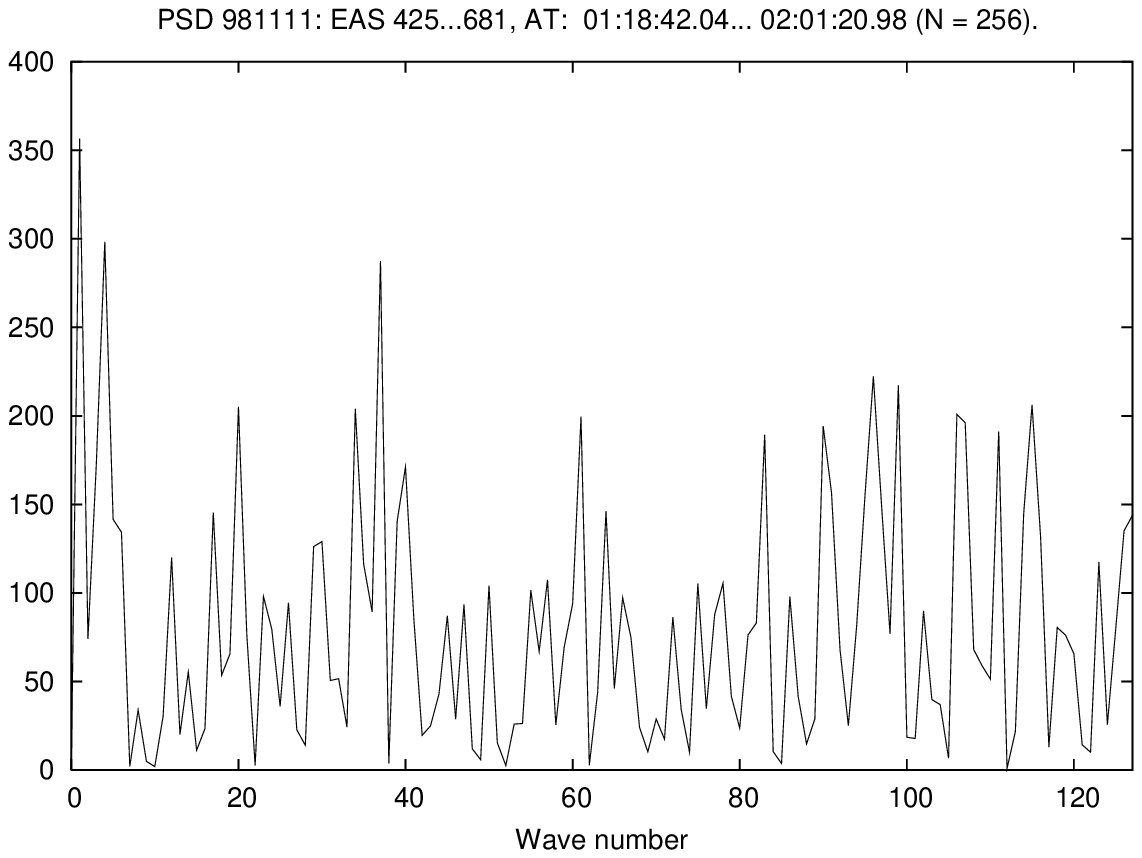}\\
	\fig{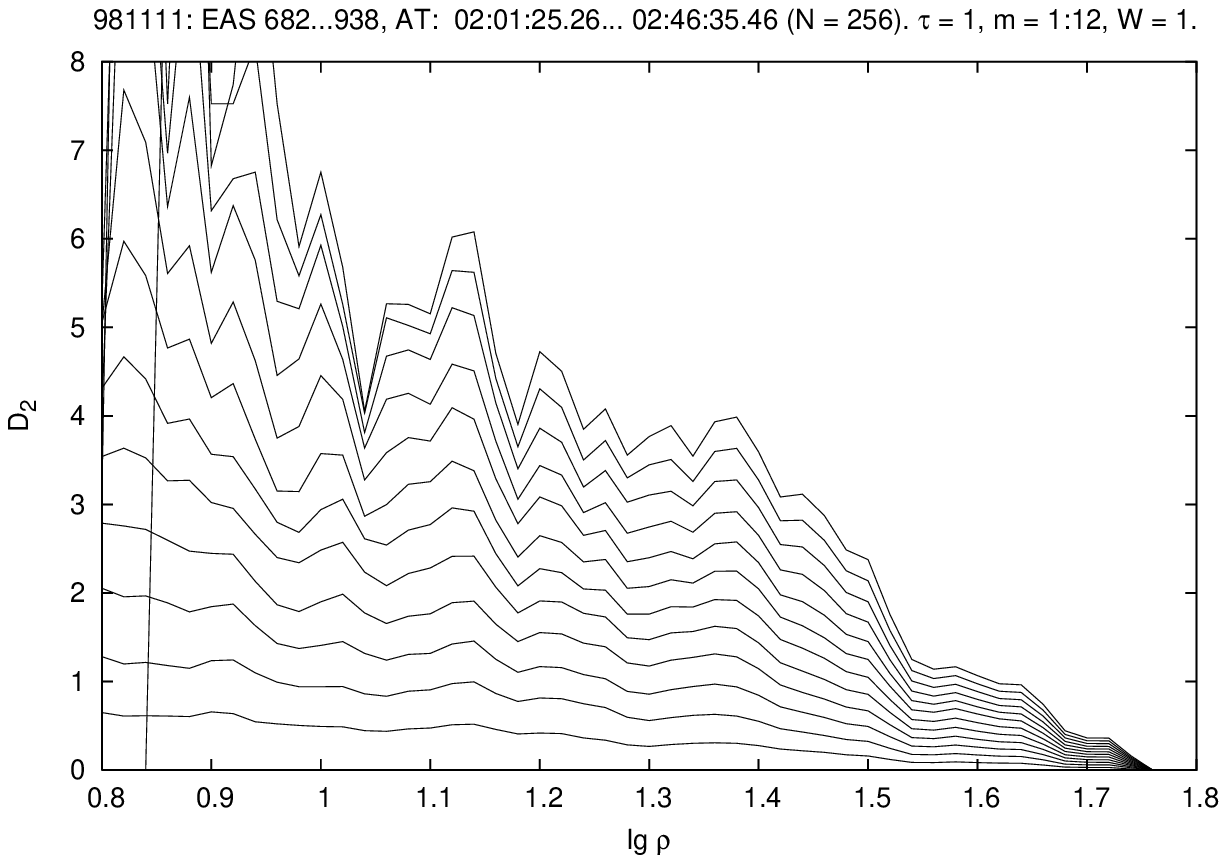}
	\fig{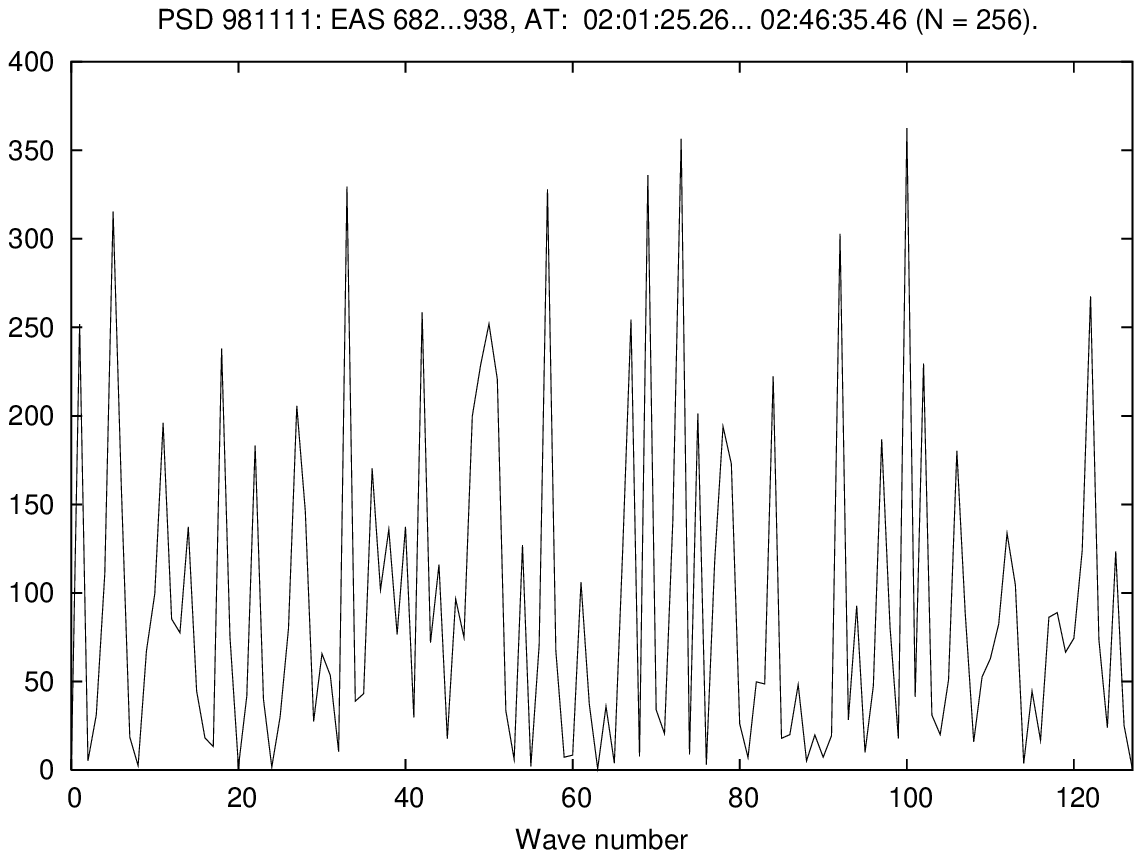}
\end{center}
\caption{The correlation dimension $D_2(\rho)$ (the left column) and
     the power spectrum density (the right column) for a number of
	samples in the vicinity of the cluster registered on November~11,
	1998.
	From top to bottom: a sample with the cluster (without the last
	delay), EAS \#435--569, $N=134$; a sample that contains the cluster
	as a subset, EAS \#425--681, $N=256$; a sample that follows the
	second one, EAS \#682--938, $N=256$. The distance~$\rho$ is given
	in seconds.}
\label{981111:D2&PSD}
\end{figure}

	Figure~\ref{981111:D2&PSD} shows the correlation dimension
	(the left column) and the power spectrum density (PSD)
	(the right one)
	calculated for three samples in the vicinity of the cluster.
	The top row represents a sample that consists of EAS \#435--569
	and thus contains the cluster without the last delay.
	The length of the sample $N=134$.
	The middle row represent a sample that contains the cluster
	but consists of $N=256$ delays (EAS \#425--681).
	The bottom row represents a sample that immediately follows
	the second one.
	It also consists of 256 delays (EAS \#682--938).
	The correlation dimension was computed for $\tau=1$, $W=1$, and
	$m=1,\ldots,12$.

	What can be seen from this figure?
	In our opinion, the most important thing is that
	one can see plateaus in both plots of~$D_2(\rho)$
	that represent samples with the cluster.
	To the contrary, no plateau is observed for the sample outside
	the cluster.
	We stress that in this sense, the sample is typical for the
	whole data set.

	Next, it is remarkable that the influence of the cluster on
	the behaviour of the correlation is not restricted to the cluster
	itself, e.g., the sample made of EAS \#425--681 begins approximately
	2.5~min before the cluster and ends up in 23.5~min after it.
	Our analysis has revealed that a plateau can be observed even for
	samples with~$N\sim500$ lying the in the range of EAS numbers
	340--1000.
	It is also interesting to mention that the plateau is more
	pronounced if a sample contains more of the after-cluster showers
	than those arrived before.

\begin{figure}
\begin{center}
	\fig{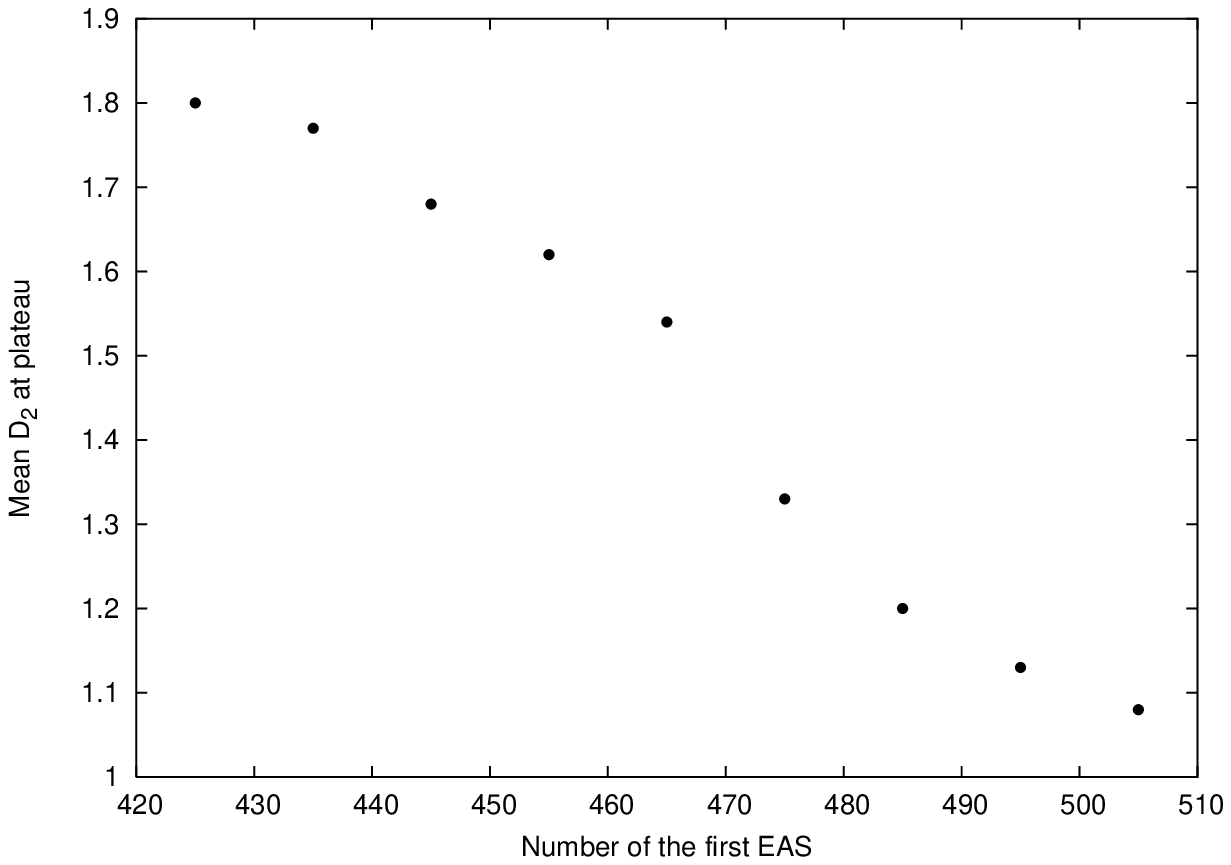}
	\fig{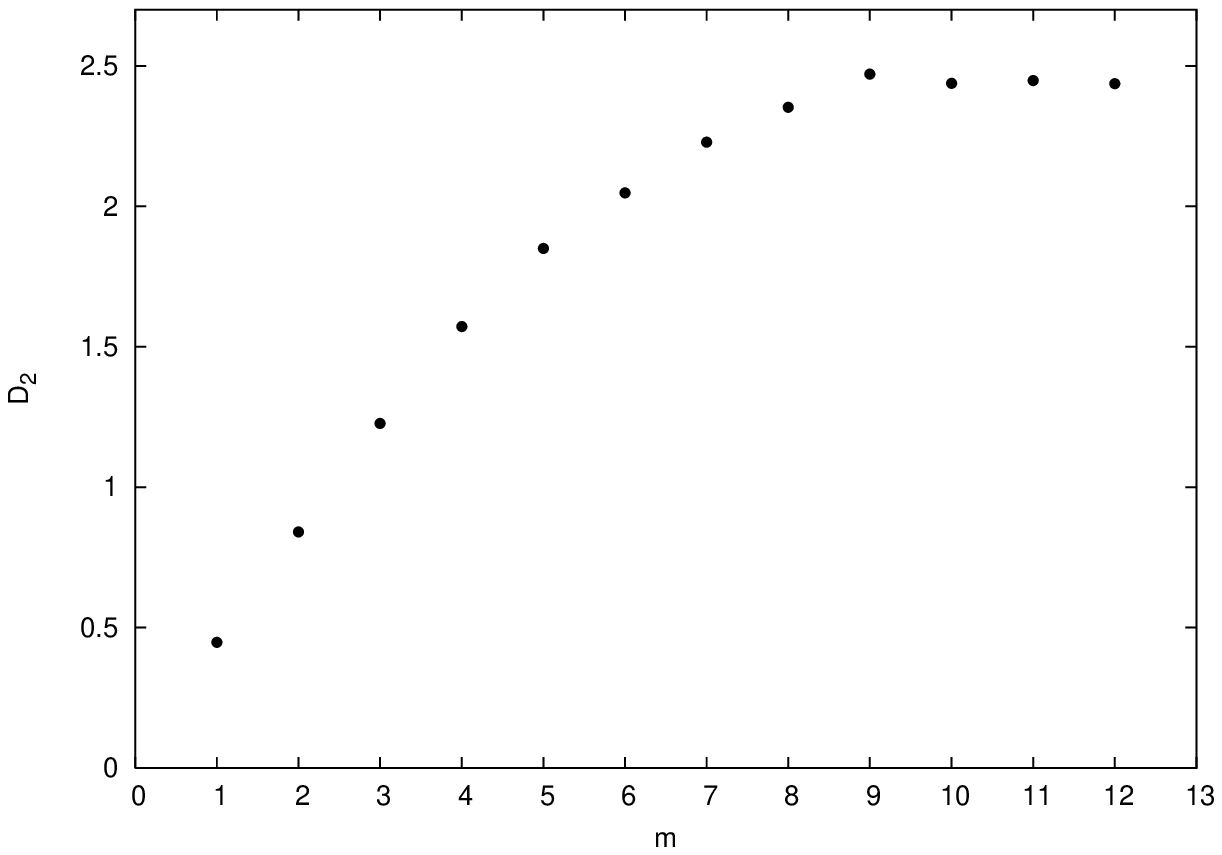}
\end{center}
\caption{
    The left panel: dependence of the correlation dimension $D_2(\rho)$
    on the position of the sample ($N=134$).
    The right panel: dependence of the correlation dimension at the
    plateau on the value of the embedding dimension
    (EAS \#425--681, $N=256$).}
\label{981111:dynamics}
\end{figure}

	The level of the plateau changes for different~$N$.
	The correlation dimension fluctuates around $D_2\approx1.77$
	for the sample with $N=134$ while $D_2\approx2.45$ for the
	sample with $N=256$.
	This may be due to low numerousness of the samples
	and/or white noise effects.\footnote{%
	We thank an anonymous referee for pointing this out to us.}
	Notice that for both samples with a plateau the demand
     $m > 2 D_2 + 1$ is satisfied if we assume that~$D_2$ is given by
     the level of the corresponding plateau (see,
     e.g.,~\cite{Theiler:JOSA}).
	We also remark here that the level of a plateau varies for
	samples with the same length but with different positions
	with respect to the cluster.
     The left panel in \Fig{981111:dynamics} shows the dependence
     of the mean value of~$D_2$ at a plateau on the position of a
     sample.
     All samples have the same length $N=134$ but begin at different
     EAS from \#425 up to \#505.
     EAS \#425 arrived at 01:18:42.04 MSK, EAS \#505 arrived at
     01:30:42.15, thus the left ends of these samples cover
     12-minute time interval.
     It is evident from the plot that the value of the correlation
     dimension takes different values and monotonically decreases 
	from $\approx1.80$ to $\approx1.08$.
     For samples that begin at EAS $\ge 510$ a plateau becomes
     shorter and soon disappears.

     The right panel in \Fig{981111:dynamics} demonstrates how
	the value of~$D_2$ saturates as the embedding dimension~$m$ grows.
     The plot is made for the sample that consists of EAS \#425--681.
	It is clearly seen that the saturation takes place at $m=9$
	but not at $m=3$ as one may expect for this value of~$D_2$ and
	a purely chaotic time series.
	As it was shown in~\cite{Ding93}, this may be due to the 
	presence of observational noise and small length of the sample.
     The demand $N\ge10^{D_2/2}$~\cite{EckmannRuelle92} is also
	obviously hold.

	Finally, one can notice that the plateaus are not as ``smooth''
	and long as one may like to see.
     In this connection, we would like to mention two things:
	(1)~Experimental data are affected by noise that can spoil the
	picture; we plan to employ special noise filtering techniques in
	future.
	(2)~We do not use any smoothing techniques while calculating~$D_2$.
	They can improve a plot of~$D_2$ but the situation remains
	qualitatively the same.

	Now let us turn to the right column of \Fig{981111:D2&PSD}.
	This column presents the results of the Fourier analysis of the
	three samples.
	The PSD of the sample without a plateau is a broadband one, which
	is typical for a random noise signal.
	The PSDs of two other samples do not also differ significantly 
	from a broadband one.
	Still it seems worth mentioning that for both samples with a
	plateau one can see the highest peaks located at the left end of
	the spectrum.
	This is true for almost all samples with a plateau in $D_2$-plots
	we have found.

     Next, let us employ the tools discussed in Section~2 in order
	to obtain a deeper insight in the observed phenomenon.
	For brevity, let us call the samples with a plateau in $D_2$-plots
	`possibly chaotic' (PC).
	Our aim is to figure out whether they are indeed chaotic or
	represent a special stochastic process.

\begin{figure}
	\begin{center}
		\fig{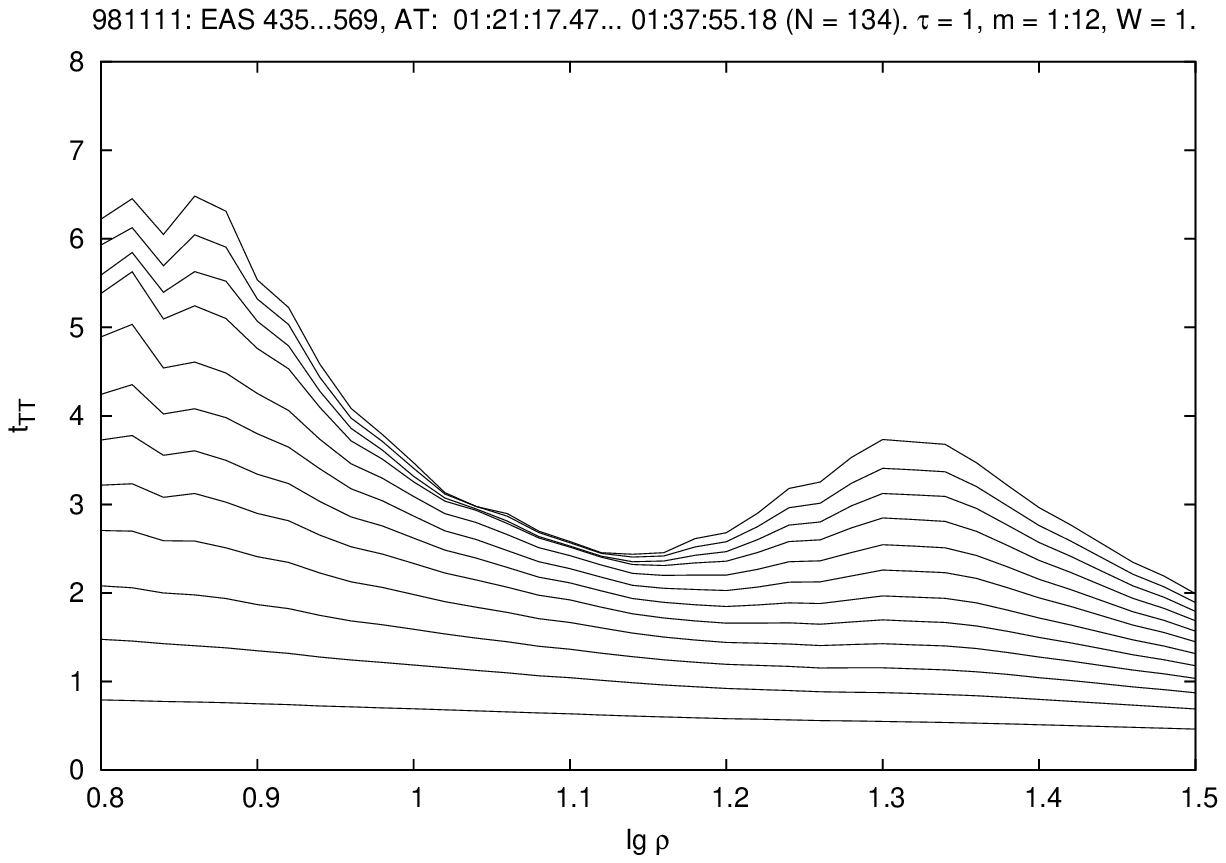}
		\fig{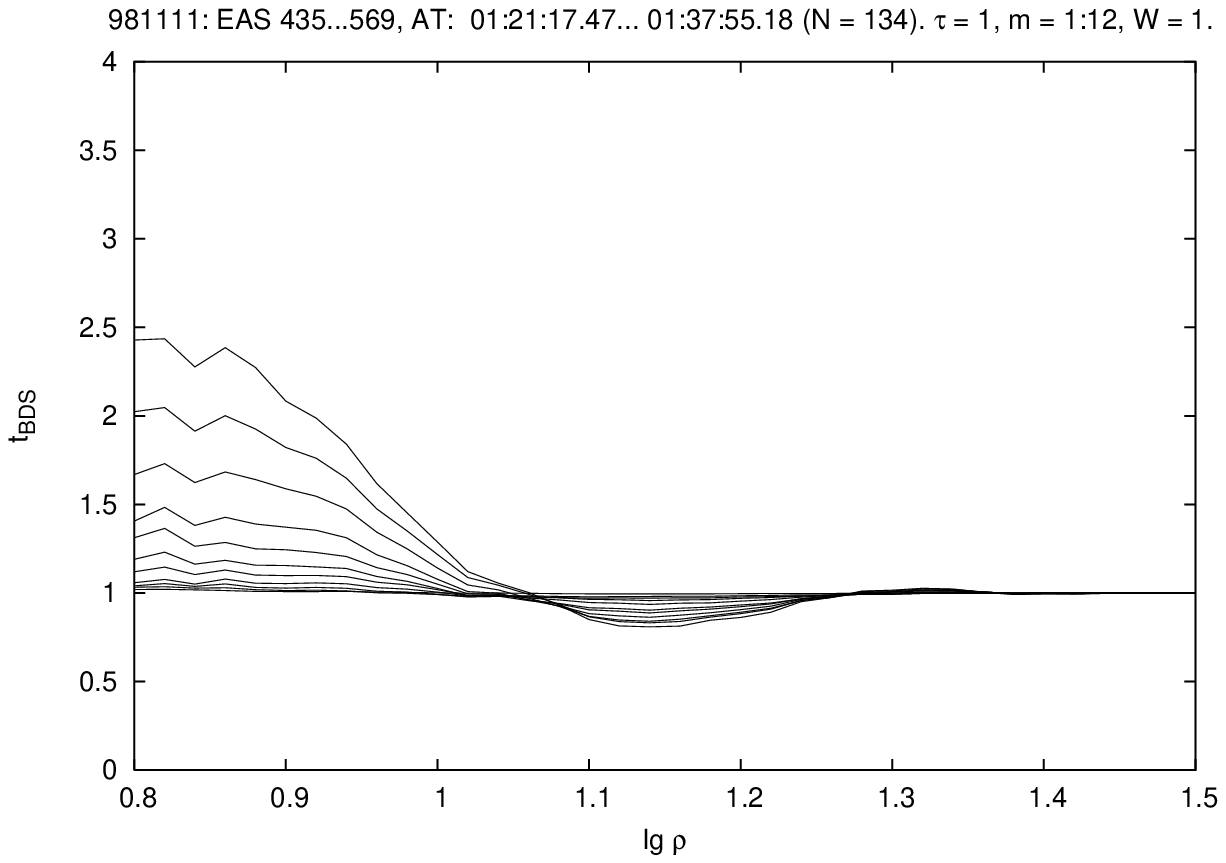}\\
		\fig{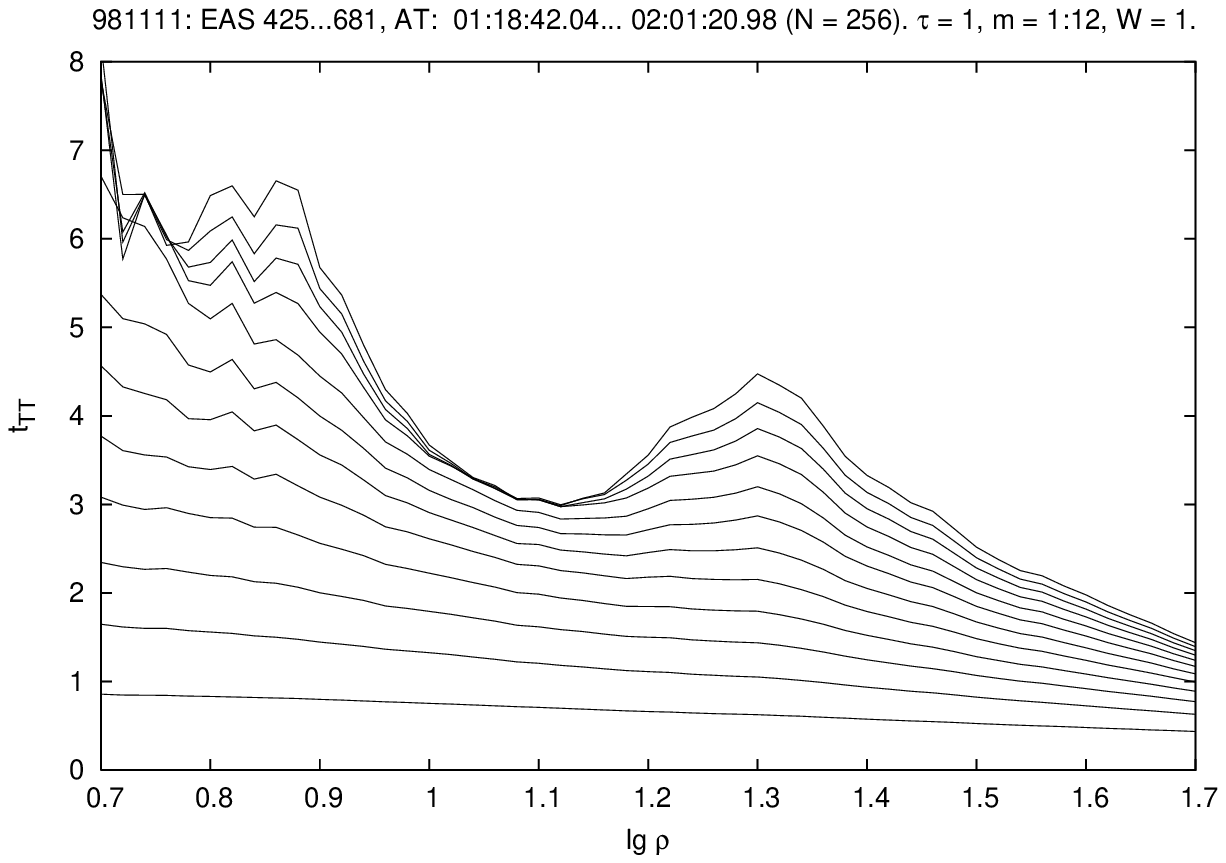}
		\fig{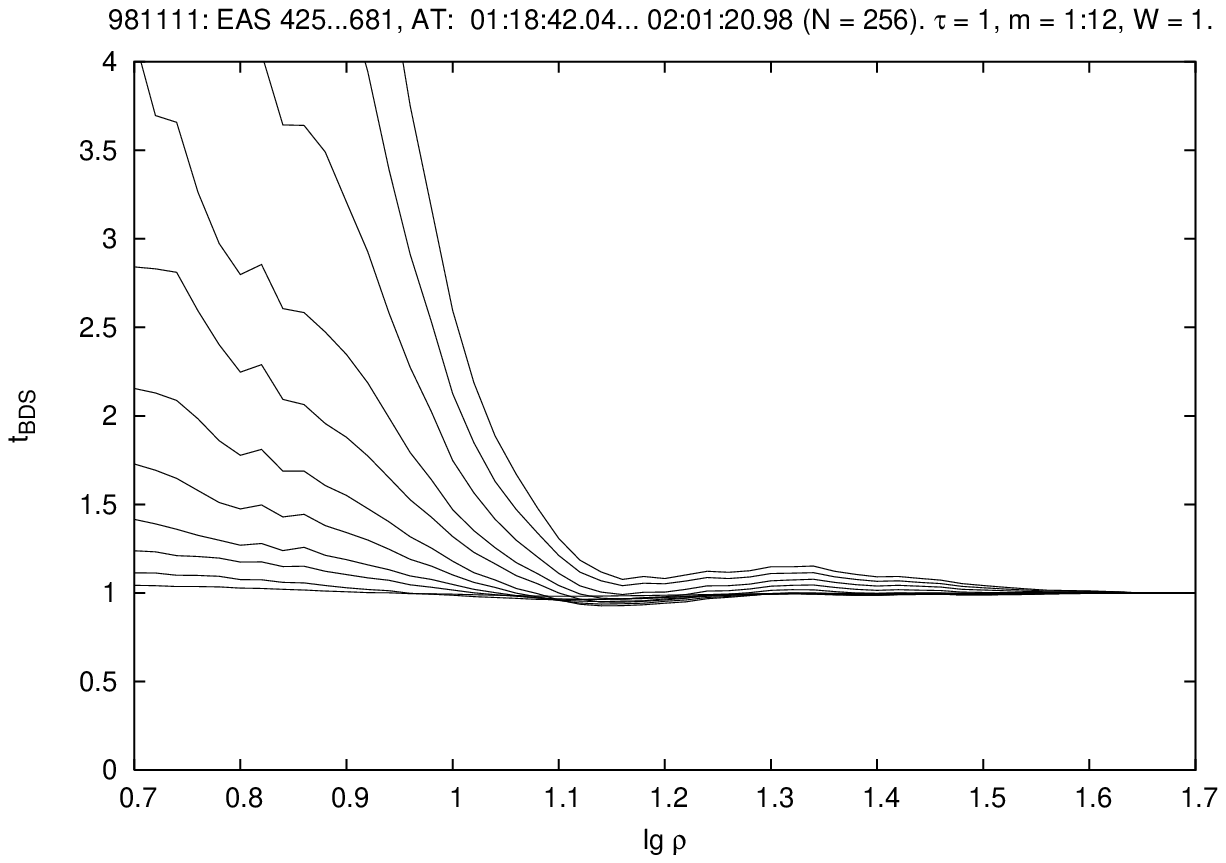}\\
		\fig{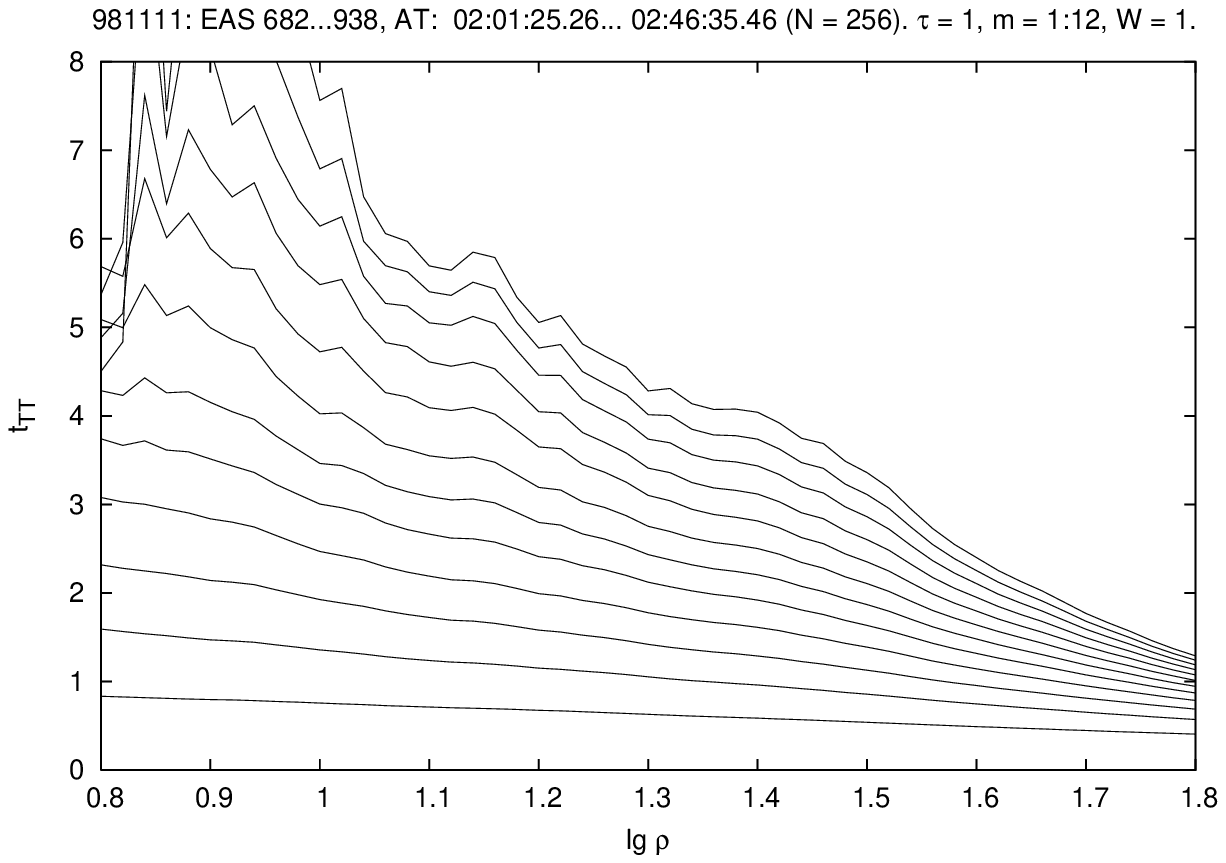}
		\fig{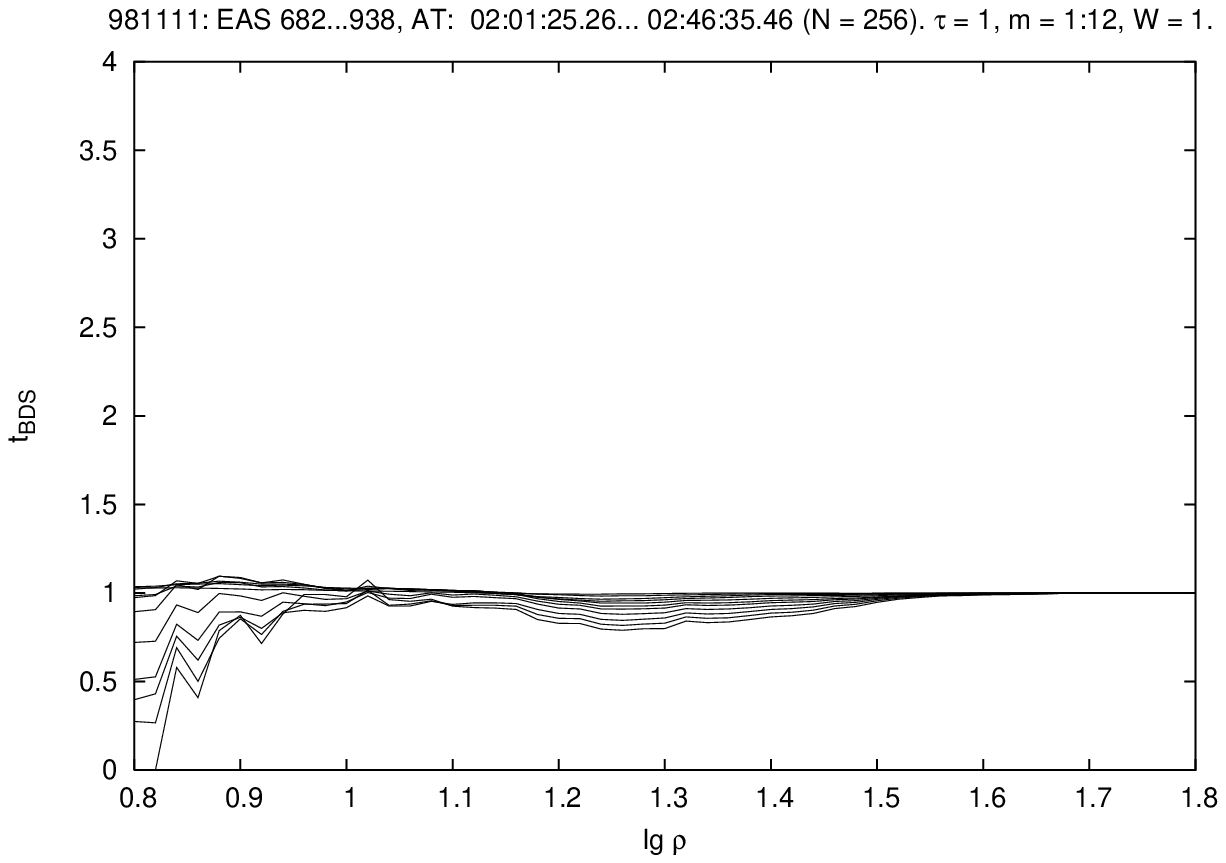}
	\end{center}
	\caption{The left column: the Theiler--Takens estimator, see
	     \Eq{eq:tTT}.
		The right column: $\tBDS$, \Eq{eq:tBDS}.
		The functions are computed for the same samples and values
		of~$\tau$, $W$, and~$m$ as in \Fig{981111:D2&PSD}.
	}
	\label{981111:tTT&tBDS}
\end{figure}

\begin{figure}
	\begin{center}
		\fig{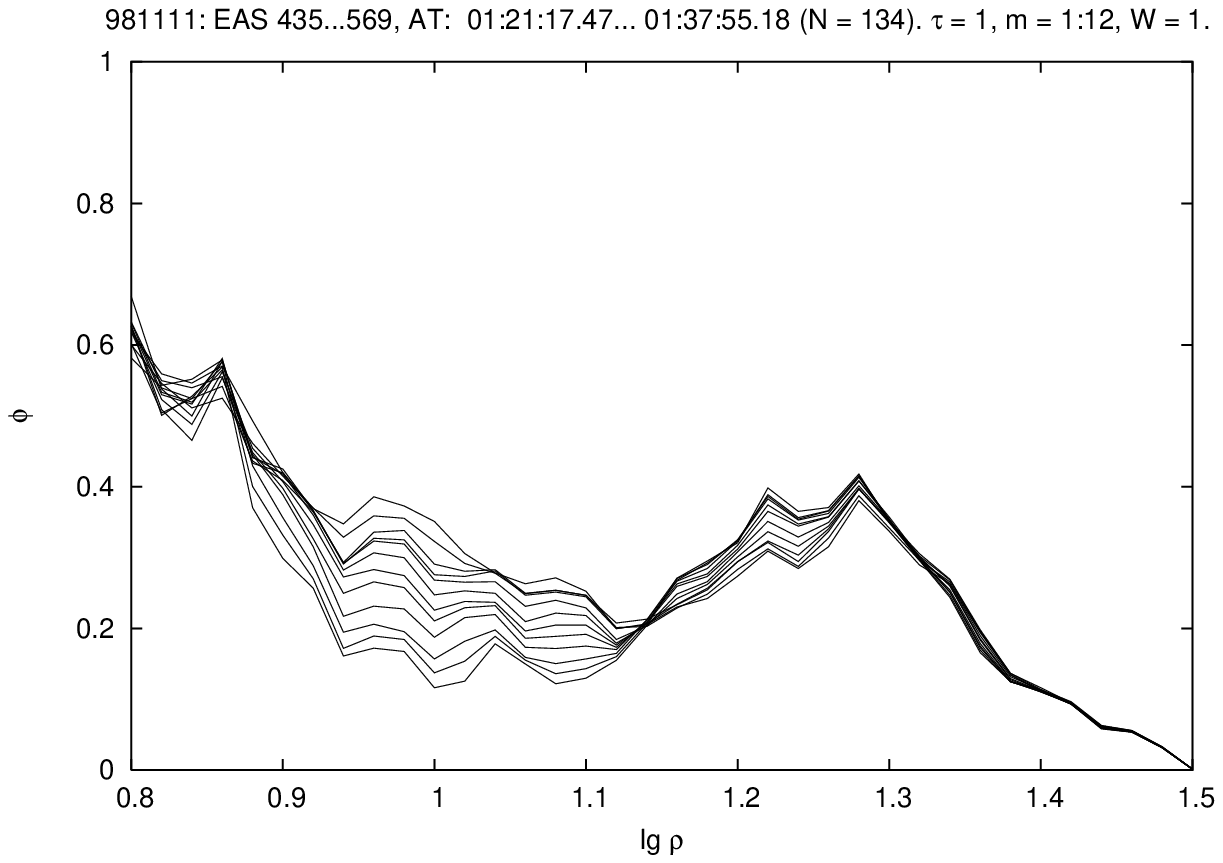}
		\fig{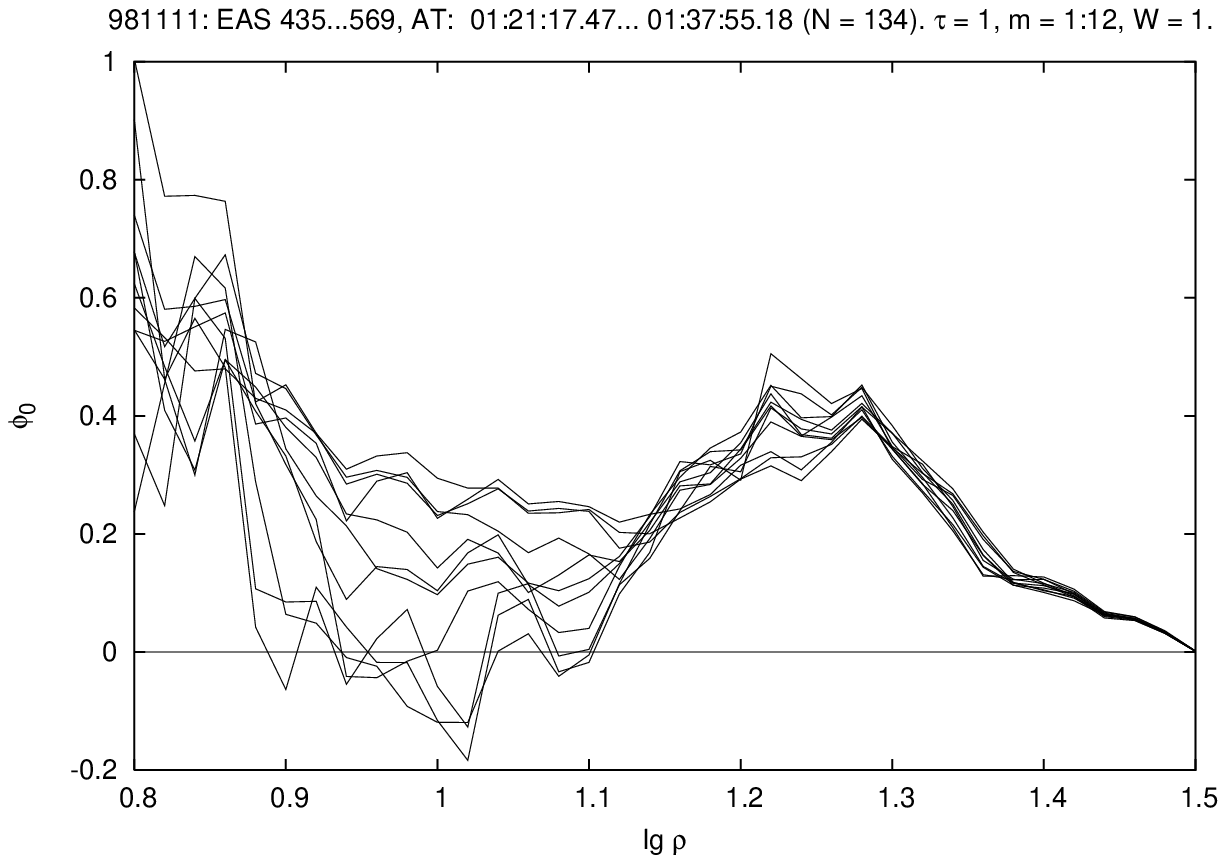}\\
		\fig{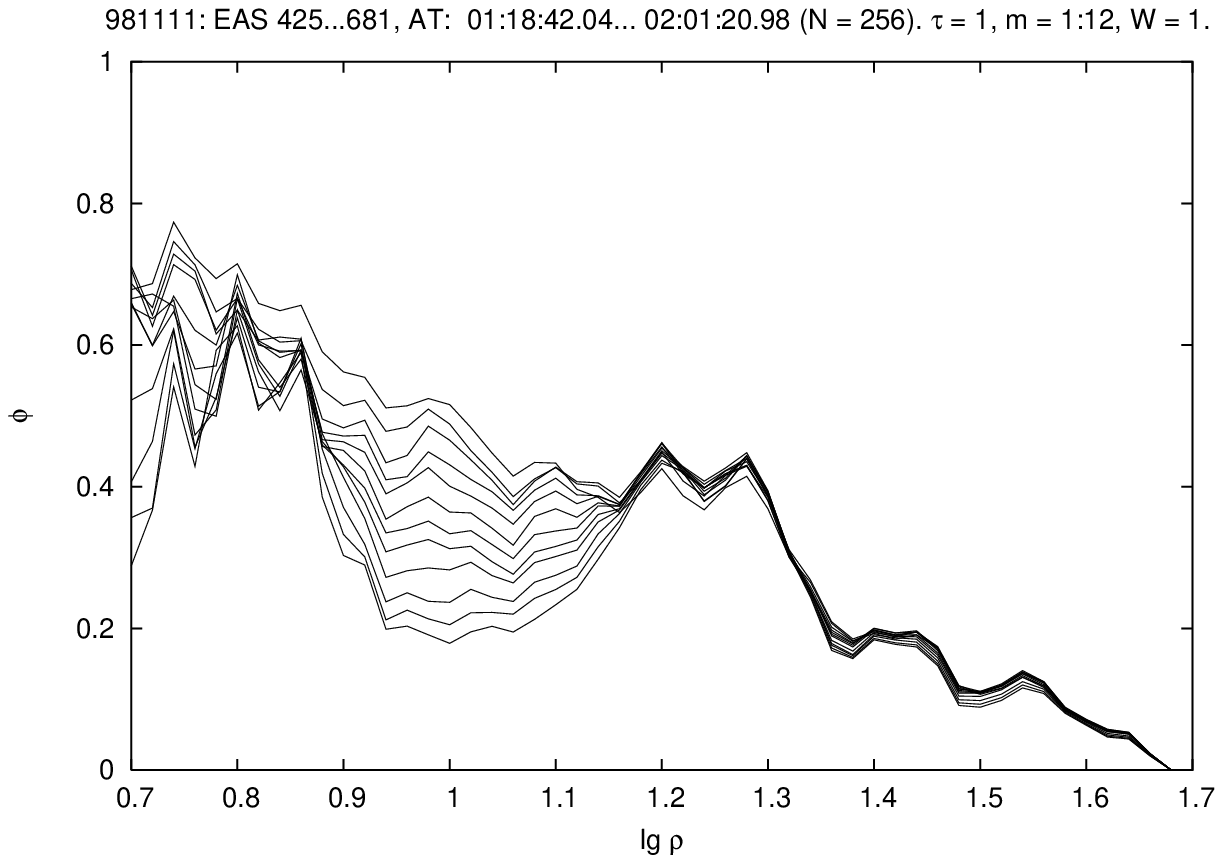}
		\fig{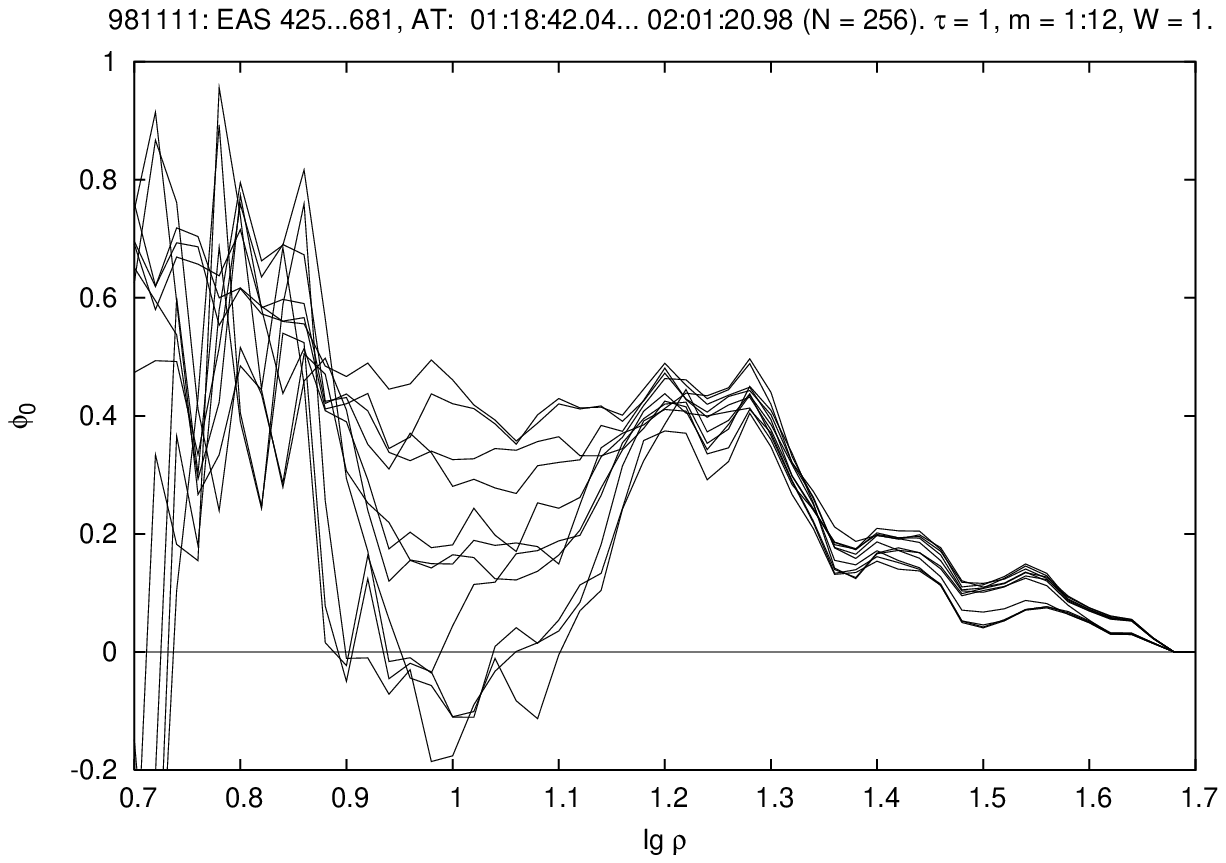}\\
		\fig{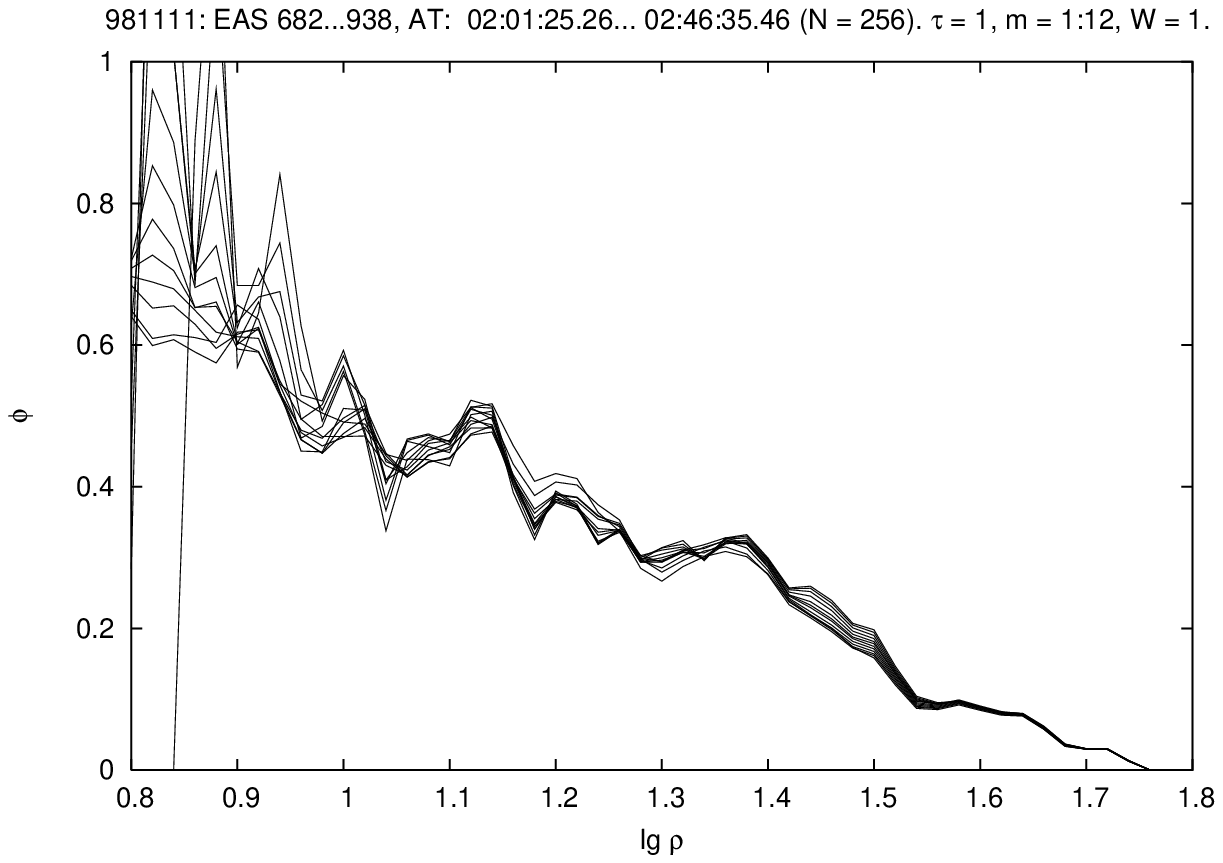}
		\fig{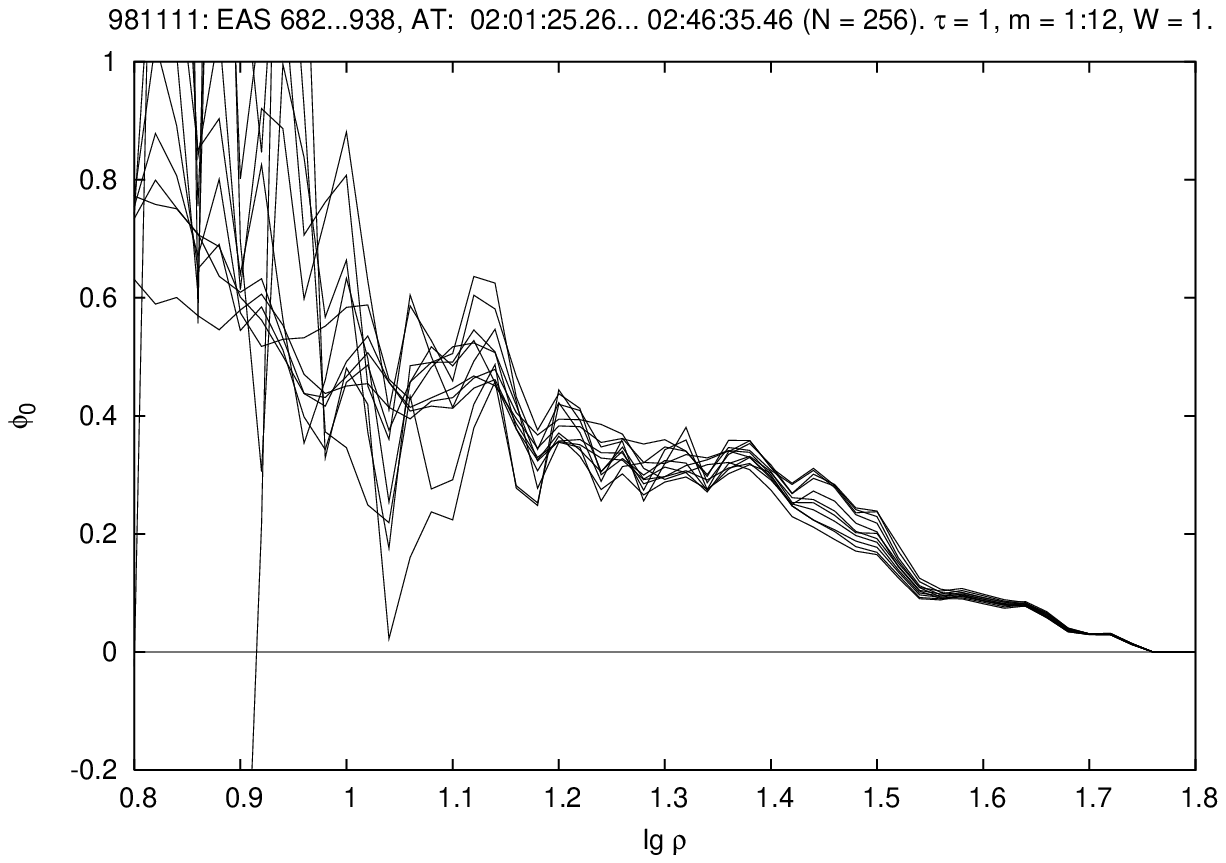}
	\end{center}
	\caption{The left column: the normalized slope~$\phi$,
     	see \Eq{eq:varphi}.
		The right column: the function~$\phi_0$ given
          by \Eq{eq:varphi0}, $m_2-m_1=1$.
		The functions are computed for the same samples and values
		of~$\tau$, $W$, and~$m$ as in \Fig{981111:D2&PSD}.
	}
	\label{981111:phis}
\end{figure}

	As one can see from \Fig{981111:tTT&tBDS}, the Theiler--Takens
	estimator~(\ref{eq:tTT}) does not have a clear plateau for either 
	samples but the behaviour of~\tTT\ for the PC samples
	noticeably differs from that for the third one.
	Namely, for the PC samples the \tTT-curves saturate for
	the corresponding intervals of $\lg\rho$ for $m>8$
	while for the third sample the curves stay separated.%
	\footnote{%
	We would like to mention that one can find samples such that
	sufficiently clear plateaus can be observed for both $D_2$-
	and \tTT-plots, see the first version of this paper
	for the details~\cite{xaoc1}.}

	Next, the right column of \Fig{981111:tTT&tBDS} depicts the
	behaviour of~\tBDS, see \Eq{eq:tBDS}.
	One can see that for both PC samples and~$m$ big enough,  \tBDS\
	noticeably differs from~1 at the intervals where plateaus of the
	correlation dimension are observed [$\lg\rho\in(0.94,1.12)$ for the
	first sample and $\lg\rho\in(0.94,1.08)$ for the second one]. 
	Conversely, $\tBDS\approx1$ for $\lg\rho\ge0.94$ for the third 
	sample.
	As it follows from the properties of~\tBDS\ discussed above,
	this kind of behaviour witnesses in favour of a chaotic nature
	of the PC samples and a stochastic nature of the sample
	without a plateau in the $D_2$-plot.

	\Fig{981111:phis} depicts the normalized slope~$\phi$
	(see \Eq{eq:varphi}) and the function~$\phi_0$
	(\Eq{eq:varphi0}) for the same three samples.
	One can see that the behaviour of these functions for the PC
	samples noticeably differs from that for the third one.
	Still, the situation is not unequivocal.
	Namely, for both PC samples, $\phi_0$ fluctuates around zero
	at the corresponding intervals of $\lg\rho$ thus giving an argument
	in favour of a chaotic nature of these samples.
     On the other hand, the limiting values of the normalized
	slope~$\phi$ (the bottom curves) lie above~0.1 at the same
	intervals of $\lg\rho$ thus suggesting a stochastic nature of the
	samples.
	This controversy can probably be resolved if we recall that~$\phi$
     converges slowly, see~\cite{PK98}.
	Really, the level of the ``hollows'' in the plots of~$\phi$
	decreases if we increase~$m$ reaching the values $\approx0.1$ and
	$\approx0.15$ for the first and the second PC samples respectively
	and $m=15$.
	For $m>15$ the number~$K$ of close delay vectors at the
	corresponding intervals becomes $\le100$ thus not providing a value
	statistically necessary  for accurate computation of the
	correlation dimension.

     Next, let us say a few words about the results of surrogate data
	tests.
	For the first PC sample,
     we generated 99 ``shuffled'' surrogates obtained by a
     random permutation of time intervals~$x_i$ in a given sample.
     This number of surrogates corresponds to a 98\%
     level of significance (L.S.) of the statistical test~\cite{SS99}.
	For the second sample, we made 39 ``shuffled'' surrogates
	(95\%~L.S.)
	and 39 surrogates based on the amplitude adjusted Fourier
     transform method proposed in~\cite{Theiler92}.
	To obtain these surrogates, we employed the TISEAN
     package~\cite{TISEAN}.
	None of these ``Fourier-based'' surrogates demonstrated a plateau
	in $D_2$-plots thus revealing that the consequence of time delays
	that constitute the original data is crucial for the appearance of
	plateaus.
	Therefore, the results of this test suggest that the PC samples
	represent a chaotic (and thus nonlinear) process.

	On the other hand, the test for time-reversibility based on
	$\gamma$-statistic~(\ref{eq:gamma}) did not detect nonlinearity
	in the PC samples.
	Namely, for the first PC sample $\gamma=0.74$ while
	$\gamma\in(-1.38,1.28)$ for the surrogates.
	For the second PC sample, $\gamma=-0.22$ while surrogates gave
	$\gamma\in(-0.44,0.45)$.
	(To compare, $\gamma=-0.38$ for the third sample.)
	These results imply that the hypothesis for time-reversibility
	of the PC samples cannot be rejected.
	Therefore, these samples may be linear (or obtained by a
	transformation of a linear process) thus excluding the possibility
	of chaotic dynamics.

	Thus, it is interesting to figure out what kind of stochastic
	processes can describe the distribution of time delays in the PC
	samples.
	To do this, we performed the $\chi^2$-test to verify
     the hypothesis that time delays between consecutive arrival times
     have an exponential distribution.
     We used different time bins in the range from~1 to 10~s with a step
     equal to 0.5~s providing that each bin taken into account contains
     more than 10~events.
     We have found that for both PC samples, this hypothesis
     should be rejected with at least 90\%~L.S. while for the third
	sample the hypothesis may be accepted with 85\%~L.S.

	In addition, we remark that for the above PC samples a plateau was
	only found for $\tau=1$ and the maximum norm.
	Still there are PC samples in the vicinity of the cluster
	that demonstrate a plateau in the $D_2$-plots for $\tau=2$
	and both ``dimension scaled'' norms, see~\cite{xaoc1} for the
	details.
	For all PC samples, a plateau is observed in a wide range
	of the Theiler window~$W$ (up to~20).

     Now let us briefly discuss three other EAS clusters that
	produce signs of chaotic behaviour in the corresponding
	time series.
	Table~1 contains certain information about these clusters
	as well as the cluster discussed above.
	Notice that while the cluster registered on January~8, 1999
	is sufficiently long ($\NEAS=134$)	two other clusters are
	short and this makes an appearance of signs of chaos
	in much longer samples very surprising.
	It is also worth mentioning that the fourth cluster
	(December~28, 1998) was registered within the period of observation
	of the GRB No.~7285 in the BATSE Catalogs~\cite{BATSE}.

\begin{table}[!ht]
\caption{EAS clusters that produce signs of chaotic dynamics}
\medskip
\begin{center}
\begin{tabular}{|c|c|c|c|c|c|}
\hline
Date    &Beginning  &End        & EAS      &\NEAS &\Padj\\
dd.mm.yy&hh:mm:ss   &hh:mm:ss   & Numbers  &      &     \\
\hline
11.11.98&01:21:17.47&01:38:02.27&435--570  &  136 &$2\times10^{-7}$\\
\hline
08.01.99&00:19:47.03&00:32:26.31&145--278  &  134 &$4\times10^{-7}$\\
\hline
14.05.98&22:24:50.09&22:24:54.70&7567--7574&   8  &$2\times10^{-8}$\\
\hline
28.12.98&15:22:33.18&15:23:17.34&5695--5712&  18  &$5\times10^{-7}$\\
\hline
\end{tabular}
\end{center}\nobreak\medskip\noindent
	Notation:
     \NEAS---the number of EAS in a cluster;
	\Padj---the probability of an appearance of the cluster
	assuming that time delays between consecutive EAS are
	adjusted to the pressure $P^*=742$~mm~Hg and
	the distribution of the number of EAS registered
	in a time unit (for the whole data set) obeys the Poisson
	distribution with the intensity $\lambda\approx5.77$~min$^{-1}$.
	Moscow local time is given.
\end{table}

	Table~2 presents the results of calculation of the correlation
	dimension for a number of samples in the vicinity of the clusters
	as well as the results of a number of tests discussed above.
	For the last three samples, the given values of~$\bar D_2$ were 
	computed for $m=15$, 17, and~14 respectively.
	Notice that the PC sample for January~8, 1999 is shifted with
	respect to the cluster.
	The cluster by itself does not produce signs of chaotic dynamics.
     For this event, a plateau can be observed not only for $\tau=1$
	but also for $\tau=2$ and~3.
	For the cluster registered on May~14, 1998, a plateau can also be
	observed for $\tau=2$.
	For the last cluster, a plateau was observed only for $\tau=1$.
	To the contrary to most other PC samples, a plateau exists not
	only for the maximum norm~$L_\infty$ but also for the 
	norms~$L_2$ and $L_{2C}$.
	For all PC samples, a plateau was found in a wide range of values
	of the Theiler window~$W$.
	This means that autocorrelation, which can lead to spurious results,
	is avoided.
	
\begin{table}[!t]
\caption{Results of different tests applied to PC samples}
\medskip
\begin{center}
\begin{tabular}{|c|c|c|c|c|c|c|c|c|c|c|}
\hline
Date    &PC EAS    &$\bar D_2$&  $\max$    &  \tBDS&$\phi$&$\phi_0$&SDT&$\gamma$-test&$\chi^2$-test\\
dd.mm.yy&Numbers   &          &$\Delta D_2$&       &      &        &   &             &             \\
\hline
11.11.98&435--569  &1.77&0.11&$\pm$&  $-$ & $+$    &$+$& $-$ & $-$ \\
        &425--681  &2.45&0.12&$\pm$&  $-$ & $+$    &$+$& $-$ & $-$ \\
\hline
08.01.99&105--232  &0.74&0.03& $+$ &  $+$ & $+$    &$+$& $-$ & $+$ \\
\hline
14.05.98&7499--7627&1.40&0.04&$\pm$&  $+$ & $+$    &$+$& $-$ & $+$ \\
\hline
28.12.98&5632--5760&1.28&0.07& $+$ &  $+$ & $+$    &$+$&$\pm$& $+$ \\
\hline
\end{tabular}
\end{center}\nobreak\medskip\noindent
	Notation: PC EAS Numbers---the range of EAS numbers for a
	PC sample  (to be compared with the range of EAS that form
	the corresponding cluster);
	$\bar D_2$---the mean value of the correlation dimension
	at a plateau for a maximum value of~$m$, see the text;
	$\max\Delta D_2$---the maximum value of the statistical
	error of~$\bar D_2$ ($\Delta D_2 (\rho)=D_2(\rho)/\sqrt{K(\rho)}$).
	Columns \tBDS, $\phi$, $\phi_0$, SDT (surrogate data test), and
	$\gamma$-test present results of the corresponding tests:
	``$+$"~if a test witnesses in favour of a chaotic nature of
	the sample under consideration, ``$-$"~otherwise; ``$\pm$"~means
	that no definite conclusion can be made.
	The last column presents the results of the $\chi^2$-test
	of the hypothesis for an exponential distribution of time delays
	within a sample: ``$+$"~means that the hypothesis may be accepted
	for some choice of time bins, ``$-$"~otherwise.
	For all data, $\tau=1$, $W=1$, and the maximum norm is used.
\end{table}

	As one can see from Table~2, the results of the tests applied to
	these three events are mostly similar to those for the first
	cluster. 
	Namely, none of the surrogate samples (39 ``shuffled'' and 39
	Fourier-based surrogates) demonstrates a plateau in the
	$D_2(\rho)$-plot thus implying a chaotic nature of the original
	samples.
	On the other hand, the test for time-reversibility usually does 
	not detect nonlinearity in the PC samples.
	The only exception is the last sample in Table~2.
	For this sample, the $\gamma$-test applied to shuffled surrogates
	gives the usual result but being applied to Fourier-based surrogates
	it rejects the hypothesis for time-reversibility with 95\%~L.S\null.
	It is worth mentioning that this result has been confirmed when
	we employed another technique of making Fourier-based surrogates
	also implemented in TISEAN.
	Next, for all PC samples the Theiler--Takens
	estimator~\tTT\ demonstrates either a plateau or an interval
	where the curves saturate as the embedding dimension~$m$
	grows.
	Finally, it is interesting to note that for three PC samples
	we have found a way of grouping time intervals that allows
	one to accept a hypothesis for an exponential distribution
	of time delays within the corresponding PC sample.
	Still we must remark that for the majority of groupings
	this hypothesis should be rejected with at least 90\%~L.S.

	Thus we come to a conclusion that the observed dynamics
	of time intervals between consecutive EAS in the vicinity
	of these four clusters may indeed be chaotic though this
	cannot be assessed unequivocally at the moment.
	Further investigation is necessary.

\section{Discussion}

\begin{figure}
	\begin{center}
          \includegraphics[width=0.7\hsize]%
          {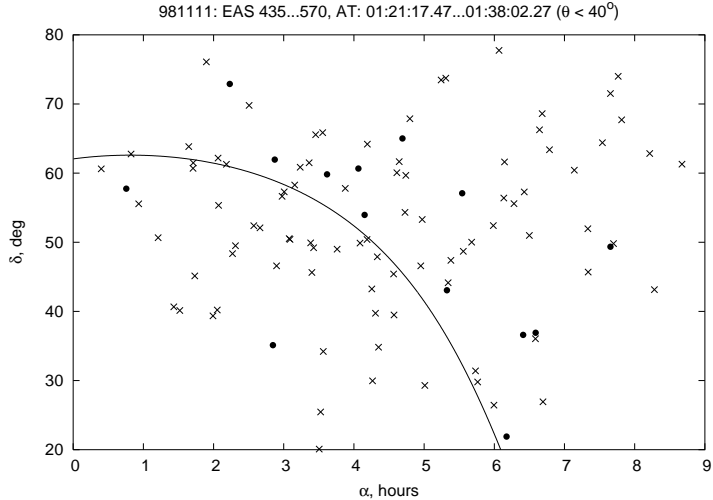}
	\end{center}
	\caption{Arrival directions (AD) of EAS that form the 
	cluster registered on November~11, 1998 and have zenith angle
	$\theta<40^\circ$: $\bullet$---AD of EAS with 
	$N_e\ge9.7\times10^4$, $\times$---AD of all other EAS.
	The curve shows the Galactic plane.
	Equatorial coordinates are used.}
	\label{981111:AD}
\end{figure}

     To get a deeper insight into physics of EAS clusters, let us
	consider the cluster registered on Nov.~11, 1998 from another
	point of view.
	A natural question that arises when one finds a burst of EAS
	count rate is whether primary particles that led to the appearance
	of the burst had had a common compact source.
	\Fig{981111:AD} shows arrival directions of EAS that belong to the
	cluster and have zenith angle $\theta<40^\circ$.
	Obviously, one cannot give a positive answer to the above question.
	Still, this seems quite natural since the energy of primary 
	particles in our case is too low to ``remember'' the source.
	Besides this, the majority of EAS in the cluster represent a
	normal count rate.
	Namely, since we register 5.7 EAS/min in average and the duration
	of the cluster adjusted to $P^*=742$~mm~Hg equals approximately
	15~min then we may expect to register 85--86 EAS within this
	period.
	Naturally, arrival directions of these EAS should be distributed 
	more or less isotropically.
	Thus, one may expect to observe a kind of a cluster in arrival
	directions for only remaining 50~EAS.
	Still, this does not happen.
	One can see doublets and triplets of showers with almost
	coincident arrival directions and a kind of grouping along the
	Galactic plane but no big groups of EAS are observed.
	
	Another question is whether clusters consist of EAS that are
	more ``powerful'' than the others.
	Surprisingly enough, but the answer is negative.
	As we have already mentioned in Section~1, the mean
	value of electrons~$\bar N_e^\mathrm{tot}$ in EAS in our data 
	set is of the order of $1.2\times10^5$ particles.
	To compare, for the above cluster $\bar N_e \sim 9.7\times10^4$,
	and only 14 EAS have~$N_e > \bar N_e$, see \Fig{981111:AD}.
	Two of these EAS have $N_e\sim10^6$.
	On the other hand, at least 41 EAS have $N_e<1/2\bar N_e$.
	(We remark that the geometry of the EAS--1000 prototype array
	does not allow one to obtain parameters of all registered EAS.)
	
     A similar situation is observed for other EAS clusters, both
	PC and ``ordinary'' ones.
	Thus, we come to the following observations: (i)~EAS in a
	cluster do not have a joint source (arrival direction);
	(ii)~the majority of EAS in a cluster have 
	$N_e<\bar N_e^\mathrm{tot}$.
	This allows us to suggest a simple conceptual model that is
	intended to explain qualitatively how EAS clusters might be
	produced.
	To do this, we need only to assume that a considerable part of
	EAS that constitute clusters are generated by charged particles.
	
	As is well known, the interstellar space is filled with strong
	and highly inhomogeneous magnetic fields.
	Still, average magnetic fields in the heliosphere are
	not strong enough to influence the dynamics of particles
	with $E\gtrsim10$~TeV/nucleon (see, e.g.,~\cite{Burger}).
	On the other hand, a series of long-term investigations based
	on data obtained with Voyager~1, 2, and other space crafts have
	demonstrated that the large-scale magnetic field strength 
	fluctuations frequently have large amplitudes and are intermittent, 
	and that regions of relatively intense magnetic fields can have 
	a radial extent of more than 10~AU, see~\cite{Burlaga-IMF} and 
	references therein.

\begin{figure}
	\begin{center}
          \includegraphics[width=0.6\hsize]{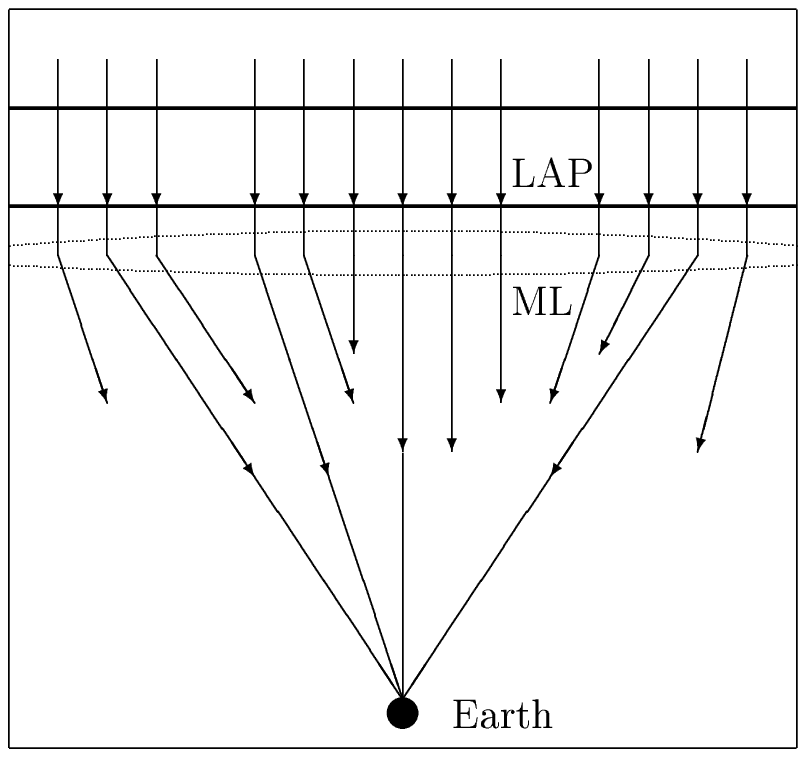}
	\end{center}
	\caption{Model of EAS clusters appearance.
	Notation: LAP---a layer of accelerated particles;
	ML---magnetic lens.}
	\label{981111:model}
\end{figure}

     Basing on these results, let us consider the following	situation.
	Suppose there is an extended and sufficiently ``thick'' layer
	of accelerated particles (LAP) (possibly called a ``wave'')
	that moves through space towards Earth, see \Fig{981111:model}.
	Normally, if it just passes through, an EAS array registers
	only a few particles originated from the LAP.
	Now suppose that the LAP meets a region of an extended strong and 
	inhomogeneous magnetic field that works as a lens.
	If it happens that this ``magnetic lens'' declines particles
	in way that they get focused in a ``proper'' direction
	then an array may register an excess of EAS over
	the normal count rate, i.e., an EAS cluster.
	Notice that from an observer's point of view showers in the
	cluster may have very different arrival directions.
	In our opinion, the fact that the majority of EAS in the observed
	clusters are less ``powerful'' than an average shower witnesses
	in favour of this model since we do not need too strong magnetic
	fields.
	The duration of a cluster depends on the thickness of the LAP
	and the time during which the magnetic lens ``works.''
	
     Now let us ask ourselves how does it happen that the time series 
	that represent the majority of EAS clusters	(as well as the other
	data) are stochastic but some clusters seem to demonstrate signs 
	of chaotic dynamics.
	At the moment, we cannot give a definite answer but can only 
	conjecture that different factors may be involved in this phenomenon.
	Among them, one can mention a possibly fractal nature of the
	interstellar medium (see, e.g.,~\cite{Lagutin} and references
	therein), nonlinear mechanisms of particles acceleration, etc.
	A very interesting possibility for EAS time series to become
	chaotic suggests the fact that the large-scale fluctuations
	in the interplanetary magnetic field strength sometimes have
	fractal or multifractal structure~\cite{Burlaga-fractals}.
	This phenomenon has been observed at different distances from
	Earth and for very different time scales.
	It is likely that in case that the ``magnetic lens'' discussed 
	above has certain fractal or multifractal properties then it
	may lead to the appearance of chaotic dynamics in EAS time series.
	
	Certainly, the presented model is only conceptual and oversimplified.
	Still we hope that it reflects the nature of processes that may
	lead to the appearance of EAS clusters, both ``ordinary'' and
	``possibly chaotic'' ones.

     Finally, it is interesting to compare our results with the 
	conclusions of similar investigations performed by other research 
	groups.
     In a considerable number of articles devoted to the nonlinear
     time series analysis, one can find a comprehensive investigation of
     EAS arrival times registered with the EAS-TOP array~\cite{Aglietta}.
     
	Basing on a detailed study of the available experimental data set
	and the results obtained with the underground muon
	monitor~\cite{Bergamasco92} the authors of this work made a
	conclusion that cosmic ray signals are all colour random noise,
	independently of the nature of the secondary particle and of the
	primary parent particle, but an existence of deterministic chaotic
	effects in cosmic ray time series cannot be completely excluded.
	It was also demonstrated in one of the following articles that
     an impact of background noise brings additional difficulties
     to the problem of distinguishing between chaotic and stochastic
     dynamics~\cite{Bergamasco94}.
	In our opinion, these conclusions as a whole do not contradict our
	results, especially in view of the fact that signs of chaotic
	behaviour have only been observed for about~0.1\% of EAS in the
	whole data set.

     Besides this, a whole series of investigations devoted to the
     nonlinear analysis of EAS time series are carried out in Japan
     beginning from early 1990s at the experimental arrays that now
     constitute the LAAS network, see, e.g., \cite{Japan:AP,Japan2001}
	and references therein.
     The authors of these investigations presented several dozens of
     events that demonstrate chaotic dynamics.
     More than this, it was conjectured that the observed dynamics
     may be due not only to the chaotic structure of the medium
     through which particles have traversed but also to the nature of the
     primary particles~\cite{Japan:particles}.
     Later on, there was suggested a model according to which chaotic
     events may be generated by cosmic rays that have a structure of a
     fractal wave arriving from a nonlinear accelerator like a supernova
     remnant~\cite{Japan:FractalWave}.
     This model needs to be studied in details, but seems to be
     promising.
	Thus that the results obtained during our analysis do not contradict 
	the conclusions of similar investigations performed at other EAS 
	arrays.
	
     The results presented above demonstrate that one can observe an
     unusual dynamics of EAS arrival times in the vicinity of certain
     clusters of EAS with the electron number of the order of~$10^5$.
	While the overwhelming majority of samples in our data set are
	unambiguously stochastic, a number of samples in the vicinity of
	four EAS clusters demonstrate signs of chaotic dynamics.
     Still it is rather difficult to make a final conclusion on the
     nature of this phenomenon: Does it represent deterministic chaos or
     a special type of a stochastic process?
     In our opinion, the majority of the tests performed witness in
     favour of the first of these two alternatives.
	Nevertheless we must mention that our analysis may somehow
	suffer of the fact that the phenomena discussed above are only
	observed at comparatively short time scales with short samples
	while time series analysis usually prefers longer samples.
     In this connection, we recall that our investigation of
     EAS clusters has revealed an existence of ``super clusters,"
     i.e., clusters that have duration more than 30~min and consist
     of hundreds of EAS.
     Our future plans include an analysis of these events.

     It seems to be necessary to continue the work in this area and
     to involve some other methods of nonlinear time series analysis.
     There are a number of other nonlinear tools that may help
     to make a more definite conclusion about the nature of the
     observed phenomenon.
     Among them, one can recall space-time--separation and recurrence
     plots and the Lyapunov exponents.
	There are also a number of other measures of nonlinearity besides
	the one used above~\cite{Schreiber:PhysRep}.
	Last but not least, special signal filtering techniques may be used
	in future to reduce effects of background noise on the dynamics
	of PC samples.

	Finally, as we have already mentioned above, perhaps the biggest
	puzzle in connection with signs of chaos in EAS time series is
	their astrophysical nature.
     It is likely that clusters that produce signs of chaotic
     dynamics in the corresponding time series are similar to the
     upper part of an iceberg in a sense that they do not present the
     complete process but only the most pronounced part of it.
     We point out that for all PC samples discussed above, the value
     of the correlation dimension is comparatively small ($D_2<3$).
     Since this value gives a lower estimate for the number of 
	degrees of freedom in the underlying process, it is possible
     that the structure of this process is not too complicated.
     Still it seems to be a great challenge to work out a good
	model that could explain chaotic dynamics in EAS arrival times.

\begin{ack}
     We gratefully acknowledge numerous useful discussions with
     A.~V.~Igoshin, A.~V.~Shirokov, and V.~P.~Sulakov who have helped
     us a lot with the experimental data set.
     We also thank Thomas Schreiber for a very helpful communication
	and anonymous referees for stimulating comments.
     This work was done with financial support of the Federal
     Scientific-Technical Program ``Research and design in
     the most important directions of science and techniques"
     for 2002--2006, contract No.\ 40.014.1.1.1110,
     and by Russian Foundation for Basic Research grant No.\
     02-02-16081.

     Only free, open source software was used for this investigation.
\end{ack}


\end{document}